\documentclass[a4paper, fleqn]{cas-dc}
\renewcommand{\printorcid}{}

\usepackage[authoryear]{natbib}
\usepackage{graphicx}
\usepackage{multirow}
\usepackage[utf8]{inputenc}
\usepackage[linesnumbered,ruled]{algorithm2e}
\usepackage{flushend}
\usepackage{subfig}
\usepackage{tikz}
\usepackage{float}

\newtheorem{definition}{Definition}
\newtheorem{example}{Example}
\newcommand\LANG{\mathord{\mathcal{L}}}

\usepackage{ulem}

\def\tsc#1{\csdef{#1}{\textsc{\lowercase{#1}}\xspace}}
\tsc{WGM}
\tsc{QE}
\tsc{EP}
\tsc{PMS}
\tsc{BEC}
\tsc{DE}

\AtBeginDocument{\setlength{\FullWidth}{\textwidth}}
\begin{document}

\let\WriteBookmarks\relax
\def\floatpagepagefraction{1}
\def\textpagefraction{.001}

\shorttitle{Planning on Discrete Event Systems Using Parallelism Maximization}
\shortauthors{L.~V.~R.~Alves, P.~N.~Pena and R.~H.~C.~Takahashi.}

\title [mode = title]{Planning on Discrete Event Systems Using Parallelism Maximization}                      

\author[1]{Lucas~V.~R.~Alves}[orcid=0000-0001-5227-0473]
\cormark[1]
\ead{lucasvra@ufmg.br}

\address[1]{Technical College, Universidade Federal de Minas Gerais}

\author[2]{ Patr\'icia~N.~Pena}[orcid=0000-0002-7595-7814]
\ead{ppena@ufmg.br}
\address[2]{Department of Electronics Engineering, Universidade Federal de  Minas Gerais}

\author[3]{ Ricardo~H.~C.~Takahashi}[orcid=0000-0003-0814-6314]
\ead{taka@ufmg.br}
\address[3]{Department of Mathematics, Universidade Federal de  Minas Gerais}

\begin{abstract}
This work deals with the production planning problem in Discrete Event Systems, using the Supervisory Control Theory to delimit the search universe and developing two heuristics based on the maximization of the parallelism to find sequences that minimize makespan. The role of the Supervisory Control Theory is to provide the set of all safe production sequences, given by the closed loop behavior. Although the use of heuristics does not provide necessarily the absolute optimal solution in general, we present a case study where it happens for all batch sizes. The efficiency in terms of computation time is also illustrated by the case study.
\end{abstract}


\maketitle

\section{Introduction}
Most of the actual industrial systems may be, at some point, modeled as a Discrete Event System (DES), where the state change is driven by the occurrence of events. In this paper, we model Discrete Event Systems using automata and languages, allowing a distinction on the system to be controlled (plant) and the controller (supervisor).
If the time required to carry out a production process in a factory is reduced, the facilities can be used to expand production and increase profit. Thus, as time is a valuable resource, choosing good operational sequences, such as those which minimize makespan, is of great importance in manufacturing. Efficient production planning using task scheduling techniques are the key to answer to the demand for efficiency \citep{4665946}.
Task scheduling refers to the allocation, over time, of finite resources to tasks in the production process, using some optimization criterion \citep{PinedoSSS}. This problem can be divided into two classes, the \textit{Deterministic Scheduling Problem} (also called \textit{model-predictive scheduling}) and the \textit{Stochastic Scheduling Problem} \citep{Aytug2005}. A scheduling is deterministic when the system is so predictable that a model can be used and the result will match the behavior of the system with negligible error \citep{SZYZZ:2007}  and, on the other hand, a scheduling is stochastic when the system is subject to unpredictable disturbances, rendering the system states predictable only in a statistical sense. There are also systems that are mostly deterministic but not completely predictable, as the deterministic schedule techniques would require \citep{Aytug2005}.
Over the years, several formalisms to address the scheduling problem emerged in the literature in the context of {\it mathematical programming} \citep{Schrijver:1986:TLI:17634, wang2019b}, {\it Petri nets} \citep{LopezMellado2005541, Wang2019, yue2016}, {\it timed automata} \citep{2006272}, {\it verification models} \citep{HerzigMBW14, Robi2018}, among others.
In the context of Supervisory Control Theory (SCT) \citep{RW89} of Discrete Event Systems (DES), there are approaches such as \citep{1678411,6042515,5771983}. Usually these techniques are focused on minimizing the makespan, the total production time of a batch of products or maximizing the throughput in the continuous production. An important advantage of such approaches is the use of the closed loop behavior of the system under SCT, that models the minimally restrictive behavior: the set of all traces that are legal. There are also constructive approaches, using {\it prioritized planning} \citep{Su2016} and  {\it sequential language projection} \citep{Ware2017,Su2017}. In common with classical deterministic scheduling solutions, approaches \citep{1678411,6042515,5771983,Su2016, Ware2017,Su2017}, consider that the duration of the operations is known and deterministic.
Finding the best sequence among all sequences of the closed loop behavior, in terms of makespan, is a non-po\-ly\-no\-mial problem \citep{1_garey_johnson_1979}, making many industrial problems intractable. To address this kind of problems the most common approaches in the literature belong to the class of heuristics \citep{Willems1994,Santos1999,Pena2016491,Almeder20112083}.
In this work, we extend the work in \citep{LucasCASE}, that proposes to minimize makespan by using a metric named parallelism maximization. As other works in the field, the closed loop behavior is used as the search universe and the duration of the operations are deterministic and known. As a development of the work in \citep{LucasCASE}, we address the temporal correctness of the resulting sequences. Instead of minimizing the makespan, we maximize the number of active parallel tasks during the production. The maximum parallel sequence is a sub-optimal solution in the time sense and this paper shows a procedure to compute such a sequence in linear time. Although the durations of the operations are known, what generates solutions with fixed positions to the uncontrollable events, we show the robustness of the approach to the case where such duration suffer random disturbances. The results show that the performance decays, but the sequence of controllable events is still valid under such variations. 

A Case study is presented, that allows to show the efficiency of the proposed approach under two aspects: i) the quality of the solution given by the algorithm in comparison to the optimal solution obtained with Model Checking \citep{Robi2018}; ii) the computation time for the solution, as long as the growth rate as the size of the batch is increased. 
This paper is structured such that in Section \ref{sec:prelim} we show some preliminary concepts and main definitions. We also show the main ideas supporting the parallel maximization. In Section \ref{sec:main1}, the two new algorithms are presented:  the algorithm of parallelism maximization taking time into consideration in  Section \ref{sec:main2} and a heuristic algorithm in Section \ref{sec:main3}.  The second algorithm is time-based but uses the idea of parallelism in the heuristic. Section \ref{sec:experim} presents the results achieved by the application of the algorithms to a manufacturing system of the literature. The conclusions and final comments are presented in Section \ref{sec:conc}.
\section{Preliminaries}\label{sec:prelim}
In this section, we summarize some fundamental concepts and results of SCT~\citep{RW89} that are needed for the development of the results presented here.
Let~$ \Sigma $ be a finite non-empty set of \textit{events}, referred to as an \textit{alphabet}.
Behaviors of DES are modeled by finite \textit{strings} over~$ \Sigma $. 
The \textit{Kleene closure} $ \Sigma^{*} $ is the set of all strings on~$ \Sigma $, including the empty string~$\varepsilon$.
A subset~$L \subseteq \Sigma^{*} $ is called a \textit{language}.
The \textit{concatenation} of strings $s,u \in \Sigma^*$ is written as~$su$.
A string~$s \in \Sigma^{*}$ is called a \textit{prefix} of $t \in \Sigma^{*}$, written $s \leq t$, if there exists $u \in \Sigma^{*}$ such that $su=t$. 
The \textit{prefix-closure} $\overline{L}$ of a language~$L \subseteq \Sigma^{*}$ is the set of all prefixes of strings in~$L$, i.e., $\overline{L} = \{\, s \in \Sigma^{*} \mid s \leq t\ \mbox{for some}\ t \in L\,\}$.
\begin{definition} \label{def:dfa}
A deterministic finite automata is a 5-tuple $G = (Q, \Sigma, \delta, q_0, Q_m)$, where $Q$ is a finite set of states, $\Sigma$ is an alphabet, $\delta : Q \times \Sigma \to Q$ is the transition function, $q_0 \in Q$ is the initial state and $Q_m \subseteq Q$ is the set of marked states.
\end{definition}
The transition function  can be  extended to recognize wor\-ds over $\Sigma^*$ as $\delta(q, \sigma s) = q'$ if $\delta(q, \sigma) = x$ and $ \delta(x, s) = q' $.
The generated and marked language are, respectively, $\mathcal{L}(G) = \{s \in \Sigma^* | \delta(q_0,s) = q' \land q' \in Q \}$ and $\mathcal{L}_m(G) = \{s \in \Sigma^* | \delta(q_0,s) = q' \land q' \in Q_m \}$.
The active event function, defined by $\Gamma : Q \to 2^{\Sigma}$, is, given a state $q$, the set of events $\sigma \in \Sigma$ for which $\delta(q,\sigma)$ is defined.
\begin{definition} \label{def:parallelcomposition}
Consider $G_1 = (Q_1, \Sigma_1, \delta_1, q_{01}, Q_{m1})$ and $G_2 = (Q_2,$ $\Sigma_2, \delta_2, q_{02}, Q_{m2})$, the synchronous product of $G_1$ and $G_2$ is:
$$G_{1||2} = (Q_1 \times Q_2, \Sigma_1 \cup \Sigma_2, \delta_{12}, (q_{01},q_{02}), Q_{m1} \times Q_{m2})$$
where
\footnotesize
\begin{equation*}
\delta((q_1, q_2), \sigma) = \begin{cases}
(\delta_1(q_1,\sigma),\delta_2(q_2,\sigma)), & \text{if $\sigma \in \Gamma_1(q_1) \cap \Gamma_2(q_2)$} \\
(\delta_1(q_1,\sigma),q_2), & \text{if $\sigma \in \Gamma_1(q_1) \backslash \Sigma_2$} \\
(q_1,\delta_2(q_2,\sigma)), & \text{if $\sigma \in \Gamma_2(q_2) \backslash \Sigma_1$} \\
\mbox{undefined}, & \text{otherwise}.
\end{cases}
\end{equation*}
\normalsize
and $\Gamma_{1||2} (q_1,q_2) = [\Gamma_1(q_1)\cap\Gamma_2(q_2)] \cup [\Gamma_1(q_1) \backslash \Sigma_2] \cup [\Gamma_2(q_2) \backslash \Sigma_1]$.
\end{definition}
The Supervisory Control Theory is a formal method, bas\-ed on the language and automata theory, to the systematic calculus of supervisors. The system to be controlled is called \textit{plant}, the controller agent is called \textit{supervisor} and the control problem is to find a supervisor which enforces the specifications in a minimally restrictive way.
The plant is modeled by an automaton $G = (Q, \Sigma, \delta, q_0, Q_m)$ and $\Sigma = \Sigma_c \cup \Sigma_u$ where $\Sigma_c$ is the set of controllable events, which can be disabled by an external agent, and $\Sigma_u$ is the set of uncontrollable events, which cannot be disabled by an external agent. The plant represents the logical model of the DES, the system behavior under no control action. The supervisor's $S$ role is to regulate the plant's  behavior to meet a desired behavior $K$ disabling controllable events.
Let $ E $ be an automaton that represents the specification imposed on $ G $.
We say that $ K = \LANG_{m}(G \parallel E) \subseteq \LANG_{m}(G) $ is \textit{controllable} with respect to $G$ if $ \overline{K} \Sigma_{uc} \cap \LANG(G) \subseteq \overline{K} $.
There exists a nonblocking supervisor $ V $ for $ G $ such that $ \LANG_{m}(V/G)= K $ if and only if $ K $ is controllable with respect to $ G $. If $ K $ does not satisfy the condition, then the \textit{supremal controllable and  nonblocking sublanguage} $ Sup \mathcal{C} (K,G) $ can be synthesized. It means that there exists a supervisor that implements such least restrictive controllable and nonblocking behavior.

The generated and marked language of a plant $G$ under the action of a supervisor $S$ are, respectively, $\mathcal{L}(S/G)$ and $\mathcal{L}_m(S/G) \subseteq \mathcal{L}(S/G)$. A supervisor $S$ is called nonblocking when $\overline{\mathcal{L}_m(S/G)} = \mathcal{L}(S/G)$. 

\subsection{Modelling Parallelism} \label{sec:parallelism}
The idea that a task is a property of a state of a deterministic finite automaton was presented in \citep{LucasCASE}. A state may have zero or more tasks being executed. If one wants only to maximize the number of machines working, the number of tasks associated to a state should be $0$ if it is an idle state or $1$ if it is a working state. If the machine has parallelism on itself, as a processor with multiple cores, the number of tasks on each state may be any non negative integer.
In order to establish the number of tasks performed in each state, we define the \textit{active task function}.
\begin{definition} \label{def:fat}
Let $G = (Q, \Sigma, \delta, q_0, Q_m)$ be a deterministic automaton. 
The \textit{active task function}, $f_{at} : Q \to \mathbb{N}$, is a function that, for every state $q \in Q$, assigns a non negative integer that represents the number of active tasks.
\end{definition}
Usually, specification automata do not perform tasks themselves.
In order to keep coherence, we may define an active task function that assigns zero tasks for all states of the specifications. 
The same maneuver should be used for plant automata which are not interesting for the optimization process.
The active task function of a composed automaton is defined as follows.
\begin{definition} \label{def:fatComposition}
Let $f_{at_{1}}$ and $f_{at_{2}}$ be the active task functions (Def. \ref{def:fat}) of $G_1$ and $G_2$, respectively. 
The active task functions of $G_{1||2} = G_1 || G_2$ is:
$$f_{at_{1||2}}(q_1,q_2) = f_{at_{1}}(q_1) + f_{at_{2}}(q_2).$$
\end{definition}
The expansion to multiple automata is straightforward.
In order to illustrate the definitions of the paper, a modified version of the \textit{small factory} \citep{W2014} is going to be used, along with the main definitions. 
\begin{example}\label{ex:Sm1} 
The small factory consists of two machines and a unity buffer, as shown in Figure~\ref{fig:sf}. The plant and specification automata for the example are presented in Fig.~\ref{fig:smaut}.
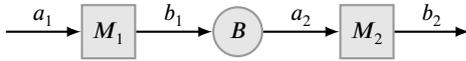
\begin{figure}[htbp]
\centering

\tikzstyle{buffer}=[circle,
                                    thick,
                                    minimum size=0.4cm,
                                    draw=gray!80,
                                    fill=gray!20]
                                    
\tikzstyle{machine}=[rectangle,
                                    thick,
                                    minimum size=0.7cm,
                                    draw=gray!80,
                                    fill=gray!20]
                                    
\tikzstyle{IO}=[rectangle,
                                    thick,
                                    minimum size=0.1cm,
                                   draw=none,fill=none]

\begin{tikzpicture}[>=latex,text height=1.5ex,text depth=0.25ex]

  \matrix[row sep=0.5cm,column sep=0.5cm] {
    &
        \node (I) [IO,draw=none]{$ $}; &
        &
        \node (m1) [machine]{$M_1$}; &
        &
        \node (b)   [buffer]{$B$};   &
        &
        \node (m2) [machine]{$M_2$}; &
        &
        \node[draw=none] (O) [IO]{$ $}; &
        \\
    };
    \path[->]
        (I)  edge[thick,left] node[above] {$a_1$} (m1)
        (m1) edge[thick,left] node[above] {$b_1$} (b)
        (b)  edge[thick,left] node[above] {$a_2$} (m2)	
        (m2) edge[thick,left] node[above] {$b_2$} (O)
        ;
\end{tikzpicture}
            
\caption{Example \ref{ex:Sm1}: Small factory diagram}
\label{fig:sf}          
\end{figure}
The active task functions $f_{at}$ for $M_1, M_2$ and $E$ are presented in TABLE~\ref{tab:tasks1}.
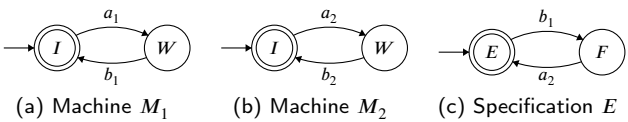
\begin{figure}[h]
\centering
\subfloat[Machine $M_1$]{
\makebox[2.5cm][c]{
\begin{tikzpicture}[scale=0.10]
\scriptsize
\tikzstyle{every node}+=[inner sep=0pt]
\draw [black] (29.5,-23.9) circle (3);
\draw (29.5,-23.9) node {$I$};
\draw [black] (29.5,-23.9) circle (2.4);
\draw [black] (44,-23.9) circle (3);
\draw (44,-23.9) node {$W$};
\draw [black] (31.992,-22.246) arc (115.73754:64.26246:10.958);
\fill [black] (41.51,-22.25) -- (41,-21.45) -- (40.57,-22.35);
\draw (36.75,-20.66) node [above] {$a_1$};
\draw [black] (41.284,-25.161) arc (-71.20944:-108.79056:14.076);
\fill [black] (32.22,-25.16) -- (32.81,-25.89) -- (33.13,-24.94);
\draw (36.75,-26.41) node [below] {$b_1$};
\draw [black] (22.5,-23.9) -- (26.5,-23.9);
\fill [black] (26.5,-23.9) -- (25.7,-23.4) -- (25.7,-24.4);
\end{tikzpicture}}
}
~
\subfloat[Machine $M_2$]{
\makebox[2.5cm][c]{
\begin{tikzpicture}[scale=0.10]
\scriptsize
\tikzstyle{every node}+=[inner sep=0pt]
\draw [black] (29.5,-23.9) circle (3);
\draw (29.5,-23.9) node {$I$};
\draw [black] (29.5,-23.9) circle (2.4);
\draw [black] (44,-23.9) circle (3);
\draw (44,-23.9) node {$W$};
\draw [black] (31.992,-22.246) arc (115.73754:64.26246:10.958);
\fill [black] (41.51,-22.25) -- (41,-21.45) -- (40.57,-22.35);
\draw (36.75,-20.66) node [above] {$a_2$};
\draw [black] (41.284,-25.161) arc (-71.20944:-108.79056:14.076);
\fill [black] (32.22,-25.16) -- (32.81,-25.89) -- (33.13,-24.94);
\draw (36.75,-26.41) node [below] {$b_2$};
\draw [black] (22.5,-23.9) -- (26.5,-23.9);
\fill [black] (26.5,-23.9) -- (25.7,-23.4) -- (25.7,-24.4);
\end{tikzpicture}}
}
~
\subfloat[Specification $E$]{
\makebox[2.5cm][c]{
\begin{tikzpicture}[scale=0.10]
\scriptsize
\tikzstyle{every node}+=[inner sep=0pt]
\draw [black] (29.5,-23.9) circle (3);
\draw (29.5,-23.9) node {$E$};
\draw [black] (29.5,-23.9) circle (2.4);
\draw [black] (44,-23.9) circle (3);
\draw (44,-23.9) node {$F$};
\draw [black] (31.992,-22.246) arc (115.73754:64.26246:10.958);
\fill [black] (41.51,-22.25) -- (41,-21.45) -- (40.57,-22.35);
\draw (36.75,-20.66) node [above] {$b_1$};
\draw [black] (41.284,-25.161) arc (-71.20944:-108.79056:14.076);
\fill [black] (32.22,-25.16) -- (32.81,-25.89) -- (33.13,-24.94);
\draw (36.75,-26.41) node [below] {$a_2$};
\draw [black] (22.5,-23.9) -- (26.5,-23.9);
\fill [black] (26.5,-23.9) -- (25.7,-23.4) -- (25.7,-24.4);
\end{tikzpicture}}
}
\caption{Example \ref{ex:Sm1}: Automata for the machines $M_1, M_2$ and the specification $E$ where $I \leftarrow$ Idle, $W \leftarrow$ Working, $E \leftarrow$ Empty, $F \leftarrow$ Full.}
\label{fig:smaut}
\end{figure}
\begin{table}[h]
\caption{Example \ref{ex:Sm1}: Tasks on each state  of components of the small factory.}
\footnotesize
\centering
\begin{tabular}{|c|c|c|}
\hline
\rowcolor[HTML]{9B9B9B} 
\begin{tabular}[c]{@{}c@{}}Automaton\end{tabular} &\begin{tabular}[c]{@{}c@{}}State ($q$)\end{tabular} & \begin{tabular}[c]{@{}c@{}}
$f_{at}(q)$
\end{tabular} \\ \hline
\multirow{2}{*}{$M_1$}&I & 0 \\ \cline{2-3}
     &W & 1 \\ \hline
\multirow{2}{*}{$M_2$}&I & 0 \\ \cline{2-3}
     &W & 1 \\ \hline
\multirow{2}{*}{$E$}&E & 0 \\ \cline{2-3}
     &F & 0 \\ \hline
     \end{tabular}
\label{tab:tasks1}
\end{table}
The composition of the two machines, $M = M_1||M_2$ is shown in Fig.~\ref{fig:sfmac} and the active task function applied over it is presented in TABLE~\ref{tab:tasksGeS}. 
\begin{table}[h]
\caption{Example~\ref{ex:Sm1}: Number of tasks of each state $q\in Q$ of the plant and of each state $y\in Y$ of the supervisor for the Small Factory.}
\footnotesize
\centering
\begin{tabular}{|c|c||c|c|}
\hline
\rowcolor[HTML]{9B9B9B} 
State ($q\in Q$) & $f_{at}(q)$ &State ($y\in Y$) & $f_{at}(y)$ \\ \hline
II & $0+0=0$& IIE & $0+0+0=0$\\ \hline
IW & $0+1=1$& WIE & $1+0+0=1$ \\ \hline
WI & $1+0=1$ & IWE & $0+1+0=1$ \\ \hline
WW & $1+1=2$ &IIF & $0+0+0=0$ \\ \hline
&&WWE & $1+1+0=2$ \\ \hline
&&IWF & $0+1+0=1$ \\ \hline
\end{tabular}
\label{tab:tasksGeS}
\end{table}
As we may see, the state $WI$ represents the machine $M_1$ in state $W$, with one active task, and machine $M_2$ on state $I$ with zero active tasks, so, the state $WI$ has one active task.
\begin{figure}[h]
\subfloat[Plant $M=(Q,\cdot,\cdot,\cdot,\cdot)$]{
\makebox[4cm][c]{
\centering
\begin{tikzpicture}[scale=0.10]
\scriptsize
\tikzstyle{every node}+=[inner sep=0pt]
\draw [black] (29.5,-23.9) circle (3);
\draw (29.5,-23.9) node {\tiny$II$};
\draw [black] (29.5,-23.9) circle (2.4);
\draw [black] (44,-23.9) circle (3);
\draw (44,-23.9) node {\tiny$WI$};
\draw [black] (29.5,-36.2) circle (3);
\draw (29.5,-36.2) node {\tiny$IW$};
\draw [black] (44,-36.2) circle (3);
\draw (44,-36.2) node {\tiny$WW$};
\draw [black] (31.992,-22.246) arc (115.73754:64.26246:10.958);
\fill [black] (41.51,-22.25) -- (41,-21.45) -- (40.57,-22.35);
\draw (36.75,-20.66) node [above] {$a_1$};
\draw [black] (41.284,-25.161) arc (-71.20944:-108.79056:14.076);
\fill [black] (32.22,-25.16) -- (32.81,-25.89) -- (33.13,-24.94);
\draw (36.75,-26.41) node [below] {$b_1$};
\draw [black] (22.5,-23.9) -- (26.5,-23.9);
\fill [black] (26.5,-23.9) -- (25.7,-23.4) -- (25.7,-24.4);
\draw [black] (27.925,-33.661) arc (-157.35416:-202.64584:9.379);
\fill [black] (27.93,-33.66) -- (28.08,-32.73) -- (27.16,-33.12);
\draw (26.7,-30.05) node [left] {$a_2$};
\draw [black] (30.86,-26.563) arc (19.01985:-19.01985:10.7);
\fill [black] (30.86,-26.56) -- (30.65,-27.48) -- (31.59,-27.16);
\draw (31.94,-30.05) node [right] {$b_2$};
\draw [black] (32.099,-34.717) arc (112.61031:67.38969:12.098);
\fill [black] (41.4,-34.72) -- (40.85,-33.95) -- (40.47,-34.87);
\draw (36.75,-33.29) node [above] {$a_1$};
\draw [black] (41.323,-37.539) arc (-69.89115:-110.10885:13.3);
\fill [black] (32.18,-37.54) -- (32.76,-38.28) -- (33.1,-37.34);
\draw (36.75,-38.85) node [below] {$b_1$};
\draw [black] (42.495,-33.618) arc (-158.5631:-201.4369:9.764);
\fill [black] (42.5,-26.48) -- (41.74,-27.04) -- (42.67,-27.41);
\draw (41.32,-30.05) node [left] {$b_2$};
\draw [black] (45.734,-26.33) arc (25.54526:-25.54526:8.627);
\fill [black] (45.73,-33.77) -- (46.53,-33.26) -- (45.63,-32.83);
\draw (47.08,-30.05) node [right] {$a_2$};
\end{tikzpicture}}

}
\subfloat[Supervisor $S=(Y,\cdot,\cdot,\cdot,\cdot)$]{
\makebox[4cm][c]{
\centering
\begin{tikzpicture}[scale=0.10]
\scriptsize
\tikzstyle{every node}+=[inner sep=0pt]
\draw [black] (29.5,-23.9) circle (3);
\draw (29.5,-23.9) node {\tiny $IIE$};
\draw [black] (29.5,-23.9) circle (2.4);
\draw [black] (42.6,-23.9) circle (3);
\draw (42.6,-23.9) node {\tiny$WIE$};
\draw [black] (42.6,-36.8) circle (3);
\draw (42.6,-36.8) node {\tiny$IIF$};
\draw [black] (29.5,-36.8) circle (3);
\draw (29.5,-36.8) node {\tiny$IWE$};
\draw [black] (29.5,-48.7) circle (3);
\draw (29.5,-48.7) node {\tiny$WWE$};
\draw [black] (42.6,-48.7) circle (3);
\draw (42.6,-48.7) node {\tiny $IWF$};
\draw [black] (22.5,-23.9) -- (26.5,-23.9);
\fill [black] (26.5,-23.9) -- (25.7,-23.4) -- (25.7,-24.4);
\draw [black] (32.5,-23.9) -- (39.6,-23.9);
\fill [black] (39.6,-23.9) -- (38.8,-23.4) -- (38.8,-24.4);
\draw (36.05,-24.4) node [below] {$a_1$};
\draw [black] (42.6,-26.9) -- (42.6,-33.8);
\fill [black] (42.6,-33.8) -- (43.1,-33) -- (42.1,-33);
\draw (42.1,-30.35) node [left] {$b_1$};
\draw [black] (39.6,-36.8) -- (32.5,-36.8);
\fill [black] (32.5,-36.8) -- (33.3,-37.3) -- (33.3,-36.3);
\draw (36.05,-36.3) node [above] {$a_2$};
\draw [black] (29.5,-33.8) -- (29.5,-26.9);
\fill [black] (29.5,-26.9) -- (29,-27.7) -- (30,-27.7);
\draw (30,-30.35) node [right] {$b_2$};
\draw [black] (29.5,-39.8) -- (29.5,-45.7);
\fill [black] (29.5,-45.7) -- (30,-44.9) -- (29,-44.9);
\draw (29,-42.75) node [left] {$a_1$};
\draw [black] (32.5,-48.7) -- (39.6,-48.7);
\fill [black] (39.6,-48.7) -- (38.8,-48.2) -- (38.8,-49.2);
\draw (36.05,-49.2) node [below] {$b_1$};
\draw [black] (42.6,-45.7) -- (42.6,-39.8);
\fill [black] (42.6,-39.8) -- (42.1,-40.6) -- (43.1,-40.6);
\draw (43.1,-42.75) node [right] {$b_2$};
\draw [black] (26.58,-48.035) arc (-108.55958:-307.12876:14.996);
\fill [black] (40.4,-21.86) -- (40.07,-20.98) -- (39.46,-21.78);
\draw (17.41,-25.66) node [left] {$b_2$};
\end{tikzpicture}
}
}\caption{Example \ref{ex:Sm1}: (a) $M$, the synchronous product of the two machines; (b) $S$, the monolithic supervisor, implementing the closed loop behaviour.}
\label{fig:sfmac}
\end{figure}
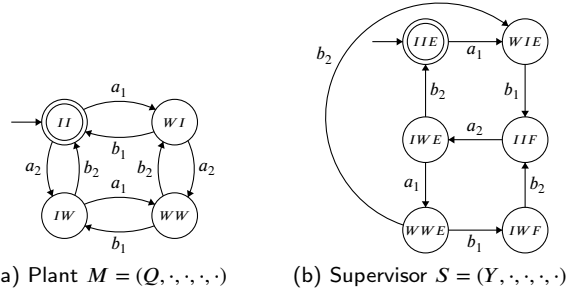
For the plant $ M $ and the specification $E$, the monolithic supervisor $S$ is presented in Fig.~\ref{fig:sfmac}(b) and the number of active tasks in each state, using Def. \ref{def:fatComposition}, is given by TABLE~\ref{tab:tasksGeS}. As the supervisor only disables events on the system, the set of states of $S$ is a subset of the set of states of $M||E$, so the states of S inherit the active task function of $M||E$.

\end{example}\vspace{.5cm}

In order to evaluate the parallelism of a string $\sigma s \in \mathcal{L}(G)$ starting on a state $q \in Q$, we define the {\it cumulative active task function}:

\begin{definition}\label{def:cumAcTaskFun}
Let $G=(Q,\Sigma,\delta,q_0,Q_m)$ be an automaton subject to $f_{at}$. The cumulative active task function $F_{at} : Q \times \Sigma^* \to \mathbb{Z}^*$ is:
\begin{align*}
F_{at} (q,\epsilon \phantom{s}) = & f_{at}(q) \\
F_{at} (q,\sigma s) = & f_{at}(q) + F_{at} (\delta(q,\sigma), s).
\end{align*}
\end{definition}

\begin{example}\label{ex:Sm31} Consider the problem of Example \ref{ex:Sm1} and two strings $s_2=a_1b_1a_2b_2a_1b_1a_2b_2$ and $s_3=a_1b_1a_2a_1b_2b_1a_2b_2$.
The cumulative active task function evaluates the sequences as $F_{at}(IIE,s_2)=4$ and
$F_{at}(IIE,s_3)=6$. Then, $s_3$ has \textit{more parallelism} than $s_2$.

\end{example}

\citep{LucasCASE} presents a polynomial algorithm that maximizes the cumulative active task function. This algorithm, referred to as the logical maximum parallelism algorithm, generates good logical solutions with very low computational cost. However, if the durations of the operations are considered, many of the solution are temporally infeasible. As mentioned in the paper, one way of mitigating such situations is to use the sequence of controllable events only and let the uncontrollable events occur as responses of the system. This approach solves the problem of temporal feasibility but, the more the real durations of the operations deviate from the ones used in the optimization, the worse becomes the solution.

In the following, we use the definitions presented before to develop two heuristics that take advantage of the SCT, of the parallelism maximization (measured by the cumulative active task function) and provide optimized solutions of great quality. The first algorithm transforms the non-polyno\-mial problem into a linear problem. The second algorithm developed from an exact algorithm, takes a heuristic step based on the idea of maximizing the cumulative active task function (Def. \ref{def:cumAcTaskFun}), providing great reduction on the branching factor. 

\section{Main Results}\label{sec:main1}
As the logical maximum parallelism algorithm, presented in \citep{LucasCASE}, uses no time information we cannot ensure that the maximum parallel sequence can be executed without any modification. This happens because the position of uncontrollable events are fixed in the solution, delayed the most to maximize parallelism, thus unlikely to execute as in the solution.

Consider the two processes of Example \ref{ex:Sm1} where the even\-ts are considered to be instantaneous but there is a delay between the controllable and uncontrollable event of each machine ($5$ $t.u.$ from $a_1$ to $b_1$ and  $10$ $t.u.$ from $a_2$ to $b_2$). When dealing with logical sequences, $a_1a_2b_2b_1\in {\cal L}_m(G)$, in Fig. \ref{fig:sfmac} a), is logically correct, part of the behavior of the plant. When taking the durations of the operations into consideration, $a_1a_2b_2b_1\in {\cal L}_m(G)$ is not temporally feasible and does not actually happen in the system. If the operations times are not taken into consideration, many temporally infeasible sequences will be generated in the optimization process and there is a considerable chance that the solution provided by the algorithm is going to be temporally infeasible.

To solve this problem, and consider only temporally feasible sequences as candidates, we extend the algorithm in \citep{LucasCASE} inserting time information in order to force the resulting sequence to be time coherent. This modification results in a non polynomial algorithm and, to keep polynomial complexity, we use a heuristic step.

\subsection{Time Constrained Maximum Parallelism}\label{sec:main2}
In order to evaluate the time until an event happens in a supervisor, given the events that already occurred, we can define a temporal function:
\begin{definition}
Let $\hat{f_T} : \Sigma^* \times \Sigma \to \mathbb{R}^*$ be the temporal function of the closed loop system $S$. Given an event $\sigma \in \Sigma$ and a sequence $s \in \mathcal{L}(S/G)$, $\hat{f_T}(s,\sigma) = t$, where $t$ is the time until the event $\sigma$ occurs given the sequence $s$ already occurred. If $\delta(\delta(q_0, s), \sigma)$ is not defined, then $\hat{f_T}(s,\sigma) = \infty$.
\end{definition}
Usually, the temporal function is implemented as an event scheduler, similar to those used on discrete event systems simulation \citep{CassandrasLafortune2ed}. Another useful measurement is the amount of time we need to execute a complete sequence of events. For this purpose we can expand the temporal function to give the time of a sequence.
\begin{definition} \label{def:temporalf}
Let $f_T : \Sigma^* \to \mathbb{R}^*$ be the extended temporal function, defined as:
\begin{align*}
f_T (\epsilon \phantom{s}) = &\; 0 \\
f_T (s \sigma) = & \;f_T(s) + \hat{f_T} (s,\sigma).
\end{align*}
\end{definition}
Next, we formulate the optimization problem under time constraints.

\subsubsection{Parallelism Maximization Problem Formulation}

Let $S$ be a supervisor for a production system $G = ||_{k=0}^{N}$ $G_k $, where  $G_k$, $k \in \{0 \ldots N\}$ is the set of subplants of the system, and let $F_{at}$ be the cumulative active task function (Def. \ref{def:cumAcTaskFun}) associated to the automaton that implements the closed loop behavior, $S/G$. Let $n$ be the number of events needed to produce a batch of products and let the search universe be the language $L = \{ s \in \mathcal{L}_m(S/G) : n = |s| \land f_T(s) \neq \infty \}$, where $f_T$ is the extended temporal function (Def. \ref{def:temporalf}) of the system. The discrete event system planning problem can be defined as an optimization problem:
$$ s^* = \underset{s \in L} {\mathrm{argmax}} \phantom{a} F_{at}(s)$$
\noindent{where $s^*$ is a sequence that maximizes the number of tasks occurring in parallel on the system and that is time correct in the temporal function $f_T$.}
\subsubsection{Algorithm}
The optimization problem can be solved as a longest path problem, where the weight of a transition is the number of active tasks on the destination state. The existence of cycles in the supervisor prevents us from using a direct approach. We must turn the automaton into an acyclic graph first, grown until the cardinality of the solution is reached.
The desired acyclic graph is given by the composition of the supervisor automaton $S$ with an unwind automaton $G_{a}$ where $\mathcal{L}_m(G_a) = \{ s \in \Sigma^* : |s|=n \}$, $\Sigma$ is the event set of $S$ and $n$ is the number of events in a batch. 

On the resulting acyclic graph, a vertex is represented as a pair $(q,k)$ were $q$ is the original state of the automaton and $k$ is the number of events occurred to reach the state $q$. Starting from the initial state $q_0$, we can travel on the graph in topological order, and so, a maximum path algorithm can be executed in linear time.

\begin{example}\label{ex:Sm3}
Consider the small factory of Example~\ref{ex:Sm1}. The unwind automaton for $n=8$ is shown in Fig.~\ref{fig:Ga}.
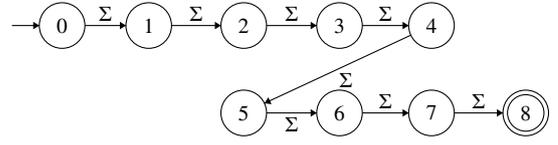
\begin{figure}
\begin{center}
\footnotesize
\begin{tikzpicture}[scale=0.10]
\tikzstyle{every node}+=[inner sep=0pt]
\draw [black] (55.6,-16.1) circle (3);
\draw (55.6,-16.1) node {$4$};
\draw [black] (18.4,-16.1) circle (3);
\draw (18.4,-16.1) node {$1$};
\draw [black] (7,-16.1) circle (3);
\draw (7,-16.1) node {$0$};
\draw [black] (30.9,-16.1) circle (3);
\draw (30.9,-16.1) node {$2$};
\draw [black] (30.9,-27.6) circle (3);
\draw (30.9,-27.6) node {$5$};
\draw [black] (43.5,-16.1) circle (3);
\draw (43.5,-16.1) node {$3$};
\draw [black] (55.6,-27.6) circle (3);
\draw (55.6,-27.6) node {$7$};
\draw [black] (43.5,-27.6) circle (3);
\draw (43.5,-27.6) node {$6$};
\draw [black] (68,-27.6) circle (3);
\draw (68,-27.6) node {$8$};
\draw [black] (68,-27.6) circle (2.4);
\draw [black] (0.4,-16.1) -- (4,-16.1);
\fill [black] (4,-16.1) -- (3.2,-15.6) -- (3.2,-16.6);
\draw [black] (10,-16.1) -- (15.4,-16.1);
\fill [black] (15.4,-16.1) -- (14.6,-15.6) -- (14.6,-16.6);
\draw (12.7,-15.6) node [above] {$\Sigma$};
\draw [black] (21.4,-16.1) -- (27.9,-16.1);
\fill [black] (27.9,-16.1) -- (27.1,-15.6) -- (27.1,-16.6);
\draw (24.65,-15.6) node [above] {$\Sigma$};
\draw [black] (33.9,-16.1) -- (40.5,-16.1);
\fill [black] (40.5,-16.1) -- (39.7,-15.6) -- (39.7,-16.6);
\draw (37.2,-15.6) node [above] {$\Sigma$};
\draw [black] (46.5,-16.1) -- (52.6,-16.1);
\fill [black] (52.6,-16.1) -- (51.8,-15.6) -- (51.8,-16.6);
\draw (49.55,-15.6) node [above] {$\Sigma$};
\draw [black] (52.88,-17.37) -- (33.62,-26.33);
\fill [black] (33.62,-26.33) -- (34.56,-26.45) -- (34.13,-25.54);
\draw (44.31,-22.36) node [below] {$\Sigma$};
\draw [black] (33.9,-27.6) -- (40.5,-27.6);
\fill [black] (40.5,-27.6) -- (39.7,-27.1) -- (39.7,-28.1);
\draw (37.2,-28.1) node [below] {$\Sigma$};
\draw [black] (46.5,-27.6) -- (52.6,-27.6);
\fill [black] (52.6,-27.6) -- (51.8,-27.1) -- (51.8,-28.1);
\draw (49.55,-27.1) node [above] {$\Sigma$};
\draw [black] (58.6,-27.6) -- (65,-27.6);
\fill [black] (65,-27.6) -- (64.2,-27.1) -- (64.2,-28.1);
\draw (61.8,-27.1) node [above] {$\Sigma$};
\end{tikzpicture}
\end{center}
\caption{Example \ref{ex:Sm3}: unwind automaton for the small factory, with a depth $n=8$ and $\Sigma=\{a_1,b_1,a_2,b_2\}$.}
\label{fig:Ga}
\end{figure}
The supervisor, when composed with an unwind automaton for the production of two products (eight events), generates the acyclic graph shown in Fig.~\ref{fig:acyclic}. Now, it is possible to apply a longest path algorithm, in order to obtain a maximum parallel sequence.
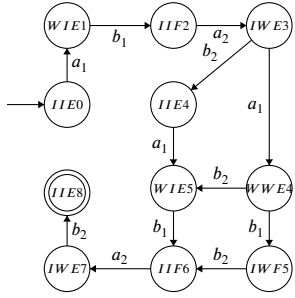
\begin{figure}
\centering
\begin{tikzpicture}[scale=0.10]
\scriptsize
\tikzstyle{every node}+=[inner sep=0pt]
\draw [black] (35.5,-23.8) circle (3);
\draw (35.5,-23.8) node {\tiny $IIE0$};
\draw [black] (35.5,-13.4) circle (3);
\draw (35.5,-13.4) node {\tiny$WIE1$};
\draw [black] (49.5,-13.4) circle (3);
\draw (49.5,-13.4) node {\tiny$IIF2$};
\draw [black] (62.1,-13.4) circle (3);
\draw (62.1,-13.4) node {\tiny$IWE3$};
\draw [black] (49.5,-23.8) circle (3);
\draw (49.5,-23.8) node {\tiny$IIE4$};
\draw [black] (62.1,-34.8) circle (3);
\draw (62.1,-34.8) node {\tiny$WWE4$};
\draw [black] (49.5,-34.8) circle (3);
\draw (49.5,-34.8) node {\tiny$WIE5$};
\draw [black] (62.1,-45.4) circle (3);
\draw (62.1,-45.4) node {\tiny$IWF5$};
\draw [black] (49.5,-45.4) circle (3);
\draw (49.5,-45.4) node {\tiny$IIF6$};
\draw [black] (35.5,-45.4) circle (3);
\draw (35.5,-45.4) node {\tiny$IWE7$};
\draw [black] (35.5,-35.4) circle (3);
\draw (35.5,-35.4) node {\tiny $IIE8$};
\draw [black] (35.5,-35.4) circle (2.4);
\draw [black] (27.6,-23.8) -- (32.5,-23.8);
\fill [black] (32.5,-23.8) -- (31.7,-23.3) -- (31.7,-24.3);
\draw [black] (35.5,-20.8) -- (35.5,-16.4);
\fill [black] (35.5,-16.4) -- (35,-17.2) -- (36,-17.2);
\draw (36,-18.6) node [right] {$a_1$};
\draw [black] (38.5,-13.4) -- (46.5,-13.4);
\fill [black] (46.5,-13.4) -- (45.7,-12.9) -- (45.7,-13.9);
\draw (42.5,-13.9) node [below] {$b_1$};
\draw [black] (52.5,-13.4) -- (59.1,-13.4);
\fill [black] (59.1,-13.4) -- (58.3,-12.9) -- (58.3,-13.9);
\draw (55.8,-13.9) node [below] {$a_2$};
\draw [black] (59.79,-15.31) -- (51.81,-21.89);
\fill [black] (51.81,-21.89) -- (52.75,-21.77) -- (52.11,-21);
\draw (54.37,-18.11) node [above] {$b_2$};
\draw [black] (62.1,-16.4) -- (62.1,-31.8);
\fill [black] (62.1,-31.8) -- (62.6,-31) -- (61.6,-31);
\draw (61.6,-24.1) node [left] {$a_1$};
\draw [black] (49.5,-26.8) -- (49.5,-31.8);
\fill [black] (49.5,-31.8) -- (50,-31) -- (49,-31);
\draw (49,-29.3) node [left] {$a_1$};
\draw [black] (62.1,-37.8) -- (62.1,-42.4);
\fill [black] (62.1,-42.4) -- (62.6,-41.6) -- (61.6,-41.6);
\draw (61.6,-40.1) node [left] {$b_1$};
\draw [black] (59.1,-45.4) -- (52.5,-45.4);
\fill [black] (52.5,-45.4) -- (53.3,-45.9) -- (53.3,-44.9);
\draw (55.8,-44.9) node [above] {$b_2$};
\draw [black] (46.5,-45.4) -- (38.5,-45.4);
\fill [black] (38.5,-45.4) -- (39.3,-45.9) -- (39.3,-44.9);
\draw (42.5,-44.9) node [above] {$a_2$};
\draw [black] (49.5,-37.8) -- (49.5,-42.4);
\fill [black] (49.5,-42.4) -- (50,-41.6) -- (49,-41.6);
\draw (49,-40.1) node [left] {$b_1$};
\draw [black] (59.1,-34.8) -- (52.5,-34.8);
\fill [black] (52.5,-34.8) -- (53.3,-35.3) -- (53.3,-34.3);
\draw (55.8,-34.3) node [above] {$b_2$};
\draw [black] (35.5,-42.4) -- (35.5,-38.4);
\fill [black] (35.5,-38.4) -- (35,-39.2) -- (36,-39.2);
\draw (36,-40.4) node [right] {$b_2$};
\end{tikzpicture}
\caption{Example~\ref{ex:Sm3}: Acyclic graph obtained by the supervisor of the Small Factory for the search depth of 8 events.}
\label{fig:acyclic}
\end{figure}

\end{example}
Find an $s^*$, when taking time into consideration, is almost as hard as finding the sequence which minimizes the makespan, so, in order to take advantage of the maximum parallelism we will use a heuristic branch and bound approach.
The inputs for the algorithm are the set of states of the supervisor ($Q$), the transition function ($\delta$), the active event function ($\Gamma$), the initial state ($q_0$) and the search depth ($depth$). As a result, the algorithm fills the structure $path$ which holds the path from the initial vertex $(q_0,0)$ to each vertex reached on the search. The composition of the supervisor with the unwind automaton is done on-the-fly during the execution of the algorithm.

\begin{algorithm}[h]

\DontPrintSemicolon
\caption{Parallelism Maximization with Time Restrictions Algorithm (PMT)}
\label{alg:PMTa}

\scriptsize
\SetKwInOut{Input}{input}\SetKwInOut{Output}{output}
\Input{$Q$, $\delta$, $\Sigma = \Sigma_{u} \cup \Sigma_{c}$, $\Gamma$, $q_0$, $depth$}
\Output{$path$}
\ForEach{state $q$ in $Q$}{
    \For{$i \leftarrow 0$ to $n$}{
        \eIf{$(q,i) = (q_0,0)$}{
            $d[(q,i)] \leftarrow 0$\;
            $path[(q,i)] \leftarrow \epsilon$\;
            $time[(q,i)] \leftarrow 0$\;
        }
        {
            $d[(q,i)] \leftarrow -\infty$\;
            $path[(q,i)] \leftarrow \emptyset$\;
            $time[(q,i)] \leftarrow \infty$\;
        }
    }
}
$F \leftarrow (q_0,0)$\;
\While{$F$ is not empty}{
    $(q,i) \leftarrow F$\;
    \lIf{$i = n$}{
        continue\;
    }
    $t_{min} \leftarrow \underset{\sigma \in \Sigma_u} {\mathrm{min}} \phantom{a} f_T(path[(q,i)] \sigma)$\;
    \eIf{$\exists \sigma \in (\Gamma(q) \cap \Sigma_u) : f_T(path[(q,i)] \sigma) < \infty$ }
     {
        \eIf{$\exists \sigma \in (\Gamma(q) \cap \Sigma_c) : f_T(path[(q,i)] \sigma) \leq t_{min}$}
        {
             $E \leftarrow \Gamma(q) \cap \{\sigma_t : \sigma_t \in \Sigma_c \land f_T(path[(q,i)] \sigma_t) \leq t_{min}\}$\;
        }
        {
            $E \leftarrow \Gamma(q) \cap \{\sigma_t : \sigma_t \in \Sigma_u \land f_T(path[(q,i)] \sigma_t) = t_{min}\}$\;
        }
     }
     {
        $E \leftarrow \Gamma(q) \cap \{\sigma_t : \sigma_t \in \Sigma \land f_T(path[(q,i)] \sigma_t) < t_{min}\}$\;
     }
    
    \ForEach{event $\sigma$ in $E$}{
        $v \leftarrow \delta(q,\sigma)$\;
        $t \leftarrow f_T(path[(q,i)] \sigma)$\;
        \lIf{$t = \infty$}{continue\;}
        \lIf{$F$ does not contain $(v, i+1)$}{
            $F \leftarrow (v, i+1)$\;
        }
        $w \leftarrow f_{ta}(v)$\;
        $d_q \leftarrow d[(q,i)]$\;
        $d_v \leftarrow d[(v,i+1)]$\;
        \If{$d_q + w > d_v$ }{ 
            $d[(v,i+1)] \leftarrow d[(q,i)] + w$\;
            $path[(v,i+1)] \leftarrow path[(q,i)] \sigma$\;
            $time[(v,i+1)] \leftarrow t$\;
        }
    }
}

\end{algorithm} 

From line 1 to 13 the structures are initialized. From line 15 to 45 a while loop is executed until the queue $F$ is empty. On line 31 the time is calculated and, on line 32, if the time to the vertex is $\infty$, the path is not timing reachable, then the vertex is ignored. 
Also knowing that, typically, the execution of a controllable event increases the number of tasks, we postpone the execution of uncontrollable events, and instead of visiting all events in $\Gamma(q)$ we only visit controllable events unless there are no controllable events active, when we visit the uncontrollable events with less time to occur (lines 19 to 28).
Lines 40 to 42 are executed if the path using event $\sigma$ is better than the previous path. As the future possible paths starting from vertex $(v,i+1)$ depend on the path from the initial vertex to $(v,i+1)$, for an exact solution we would have to keep all paths to $(v,i+1)$ because, maybe, they are not good at this point but may be far better in the future. In order to maintain the algorithm polynomial in complexity, we take a greedy step and keep only one of the best paths.
It is important to note that in order to ensure that the algorithm reaches a solution, we have to limit the number of times each controllable event may occur. This makes sense since we know the size of the batch we intend to produce and the recipe to produce it. When there are paths with the same size that produce different products, instead of using depth, other stop criteria may be used, such as the a number of occurrences of some event.
The complexity of the algorithm is the same of a breadth-first search, $\mathcal{O}(v + a)$ where $v$ is the number of vertices and $a$ is the number of edges. In this algorithm, a vertex is a state in determined depth, so for a depth of $n$ events, the number of vertices is $v = (n+1)|Q|$ and the number of edges is $a = n|\rightarrow|$, where $Q$ and $\rightarrow$ are, respectively, the set of states and the set of transitions of the supervisor, so the complexity is, in the worst case scenario, $\mathcal{O}(n|W| + n|\rightarrow|)$.

A detailed example of the execution of the algorithm is presented in the Appendix.

\subsection{Makespan Minimization Heuristic Solution}\label{sec:main3}

Following the idea of parallelism maximization and that controllable events should increase the number of tasks being executed, we propose a time-oriented heuristic that consists in applying the same delay of uncontrollable events used in Algorithm~\ref{alg:PMTa}, which seems to increase parallelism and reduce the branch-factor, but instead of maximizing parallelism, we minimize the makespan.
\subsubsection{Makespan Minimization Problem Formulation}

Let $S$ be a supervisor for a production system $G = ||_{k=0}^{N}$ $G_k $, where  $G_k$, $k \in \{0 \ldots N\}$ is the set of sub-plants of the system, and let $f_{T}$ be the temporal function (Def. \ref{def:temporalf}) associated to the automaton that implements the closed loop behavior, $S/G$. Let $n$ be the number of events needed to produce a batch of products and let the search universe be the language $L = \{ s \in \mathcal{L}_m(S/G) : n = |s| \land f_T(s) \neq \infty \}$. The discrete event system planning problem can be defined as an optimization problem:
$$ s^* = \underset{s \in L} {\mathrm{argmin}} \phantom{a} f_{T}(s)$$
\noindent{where $s^*$ is a sequence that minimizes the makespan of the production batch.}
\subsubsection{Algorithm}
The algorithm follows the same logic of an exact algorithm. A state, when reached by different paths, with different schedulers, is kept duplicated to the next iteration. A path is only discarded when there is another path that reaches the same state with smaller makespan. As in Algorithm~\ref{alg:PMTa}, for the algorithm to reach a solution, we have to limit the number of occurrences of each controllable event.

To execute the algorithm, the state set of the supervisor ($Q$), the event set of the supervisor ($\Sigma = \Sigma_{u} \cup \Sigma_{c}$), the transition function ($\delta$), the active event function ($\Gamma$), the initial state ($q_0$) and the search depth ($n$) are necessary. The structure $a$ is an event scheduler such that $a[\sigma]$ in a state $(q, i)$ is equivalent to $f_T(path[(q,i)] \sigma)$. Again, the closed loop behavior automaton should be composed with a unwind automaton, but this operation is performed on-the-fly within the algorithm.

\begin{algorithm}[ht!]
\small
\DontPrintSemicolon
 \caption{Heuristic Makespan Minimization Algorithm (HMM)}
 \label{alg:HMM}
 \scriptsize
\SetKwInOut{Input}{input}\SetKwInOut{Output}{output}
\Input{$Q$, $\delta$,$\Sigma = \Sigma_{u} \cup \Sigma_{c}$, $\Gamma$, $q_0$, $depth$}
\Output{$path$, $time$}

 $path[(q,a,0)] \leftarrow \epsilon$\;
 $time[(q,a,0)] \leftarrow 0$\;

 \;
 $F \leftarrow (q_0,a,0)$\;
 \;
 \While{$F$ is not empty}
 {
     $(q,a,i) \leftarrow F$\;
    
      \eIf{$\exists \sigma \in (\Gamma(q) \cap \Sigma_u) : f_T(path[(q,a,i)] \sigma) < \infty$ }
     {
        $t_{min} \leftarrow \underset{\sigma \in \Sigma_u} {\mathrm{min}} \phantom{a} f_T(path[(q,a,i)] \sigma)$\;
        \eIf{$\exists \sigma \in (\Gamma(q) \cap \Sigma_c) : f_T(path[(q,a,i)] \sigma) \leq t_{min}$}
        {
             $E \leftarrow \Gamma(q) \cap \{\sigma_t : \sigma_t \in \Sigma_c \land f_T(path[(q,a,i)] \sigma_t) \leq t_{min}\}$\;
        }
        {
            $E \leftarrow \Gamma(q) \cap \{\sigma_t : \sigma_t \in \Sigma_u \land f_T(path[(q,a,i)] \sigma_t) = t_{min}\}$\;
        }
     }
     {
        $E \leftarrow \Gamma(q) \cap \{\sigma_t : \sigma_t \in \Sigma\}$\;
     }
    
     \ForEach{event $\sigma$ in $E$}
     {
         $v \leftarrow \delta(q,\sigma)$\;
         $t \leftarrow f_T(path[(q,a,i)] \sigma)$\;
         $a_n \leftarrow update(a,\sigma)$\;
        
         \lIf{$t = \infty$}{continue\;}
        
         \If{$F$ does not contain $(v, a_n, i+1)$}{
             $F \leftarrow (v, a_n, i+1)$\;
         }
        
         $t_{tot} \leftarrow time[(q,a,i)] + t$\;
        
         \If{$(\nexists \phantom{a} time[(v,a_n,i+1)])$ OR $(t_{tot} < time[(v,a_n,i+1)])$}{
             $path[(v,a_n,i+1)] \leftarrow path[(q,a,i)] \sigma$\;
             $time[(v,a_n,i+1)] \leftarrow t_{tot}$\;
         }
     }
 }
 \end{algorithm}

The algorithm initializes the $path$ to the initial state as the empty sequence ($path[(q_0,a_0,0)] \leftarrow \epsilon$) and the initial time as zero ($time[(q_0,a_0,0)] \leftarrow 0$). It is important to note that in this algorithm a vertex has the form $(q,a,i)$ where $q$ is an state of the supervisor, $a$ is the event schedule and $i$ is the depth.

The initial state is inserted into the queue $F$ and, while $F$ is nonempty, the first item is removed from the queue. The heuristic part consists in giving priority to the execution of controllable events over the uncontrollable events. The algorithm verifies if there are transitions triggered by controllable events and if these transitions do not increase the timer (the time before the occurrence of the event is equal to the time after the occurrence of the event). If these transitions exist, the events which trigger them are inserted in the set of events to be evaluated ($E$). If there are no controllable transitions in the set of events to be evaluated or if they increase the timer, then only the events (controllable or not) which cause the smallest increase in time are executed ($t_{min}$).

For each event in the set $E$, the algorithm calculates the increase in time for the transition, updates the event scheduler and verifies if the obtained sequence is temporally feasible. If the destination state was not evaluated yet, it is inserted in the queue $F$. When the transitions lead to a state with a shorter production time, the path is taken as the best to that state. As a vertex visited by the algorithm is represented by $(q,a,i)$, two vertices with the same state $q$ in the same depth $i$ are treated as different vertices when they have different event schedulers.

The Heuristic Makespan Minimization (HMM) Alg\-ori\-thm  presents non-polynomial complexity because it duplicates states when the path converges to the same state with different event schedules, but the heuristic reduces the branching factor, allowing reasonable run-times even in large problems. 

Next we present a case study, in which we compare the two algorithms one with another and with results presented by \cite{Robi2018}, and analyze the quality, in terms of makespan, of the sequence that is returned.
\section{Case Study}
\label{sec:experim}

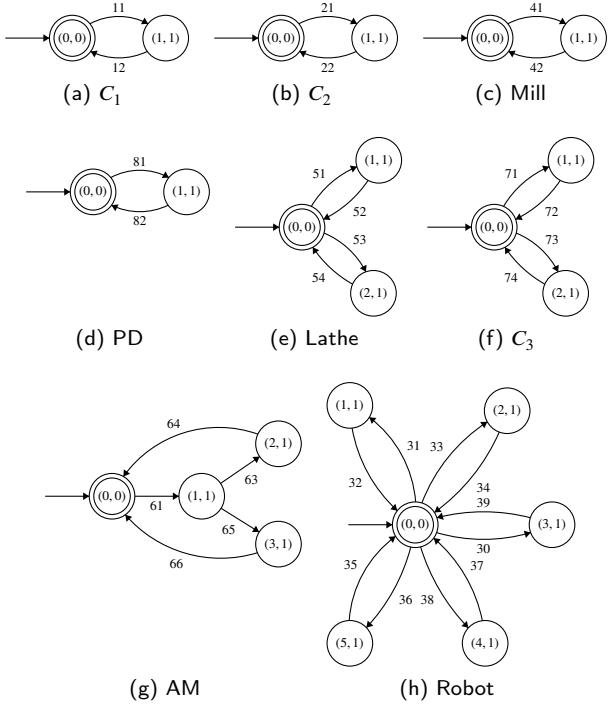
\begin{figure}
\centering

\subfloat[$C_1$]{
\begin{tikzpicture}[scale=0.10]
\tiny
\tikzstyle{every node}+=[inner sep=0pt]
\draw [black] (32.5,-27) circle (3);
\draw (32.5,-27) node {$(0,0)$};
\draw [black] (32.5,-27) circle (2.4);
\draw [black] (44.8,-27) circle (3);
\draw (44.8,-27) node {$(1,1)$};
\draw [black] (23.7,-27) -- (29.5,-27);
\fill [black] (29.5,-27) -- (28.7,-26.5) -- (28.7,-27.5);
\draw [black] (34.818,-25.123) arc (118.34915:61.65085:8.07);
\fill [black] (42.48,-25.12) -- (42.02,-24.3) -- (41.54,-25.18);
\draw (38.65,-23.66) node [above] {$11$};
\draw [black] (42.382,-28.75) arc (-64.14001:-115.85999:8.557);
\fill [black] (34.92,-28.75) -- (35.42,-29.55) -- (35.86,-28.65);
\draw (38.65,-30.11) node [below] {$12$};
\end{tikzpicture}} 
~
\subfloat[$C_2$]{
\begin{tikzpicture}[scale=0.10]
\tiny
\tikzstyle{every node}+=[inner sep=0pt]
\draw [black] (32.5,-27) circle (3);
\draw (32.5,-27) node {$(0,0)$};
\draw [black] (32.5,-27) circle (2.4);
\draw [black] (44.8,-27) circle (3);
\draw (44.8,-27) node {$(1,1)$};
\draw [black] (23.7,-27) -- (29.5,-27);
\fill [black] (29.5,-27) -- (28.7,-26.5) -- (28.7,-27.5);
\draw [black] (34.818,-25.123) arc (118.34915:61.65085:8.07);
\fill [black] (42.48,-25.12) -- (42.02,-24.3) -- (41.54,-25.18);
\draw (38.65,-23.66) node [above] {$21$};
\draw [black] (42.382,-28.75) arc (-64.14001:-115.85999:8.557);
\fill [black] (34.92,-28.75) -- (35.42,-29.55) -- (35.86,-28.65);
\draw (38.65,-30.11) node [below] {$22$};
\end{tikzpicture}}
~
\subfloat[Mill]{
\begin{tikzpicture}[scale=0.10]
\tiny
\tikzstyle{every node}+=[inner sep=0pt]
\draw [black] (32.5,-27) circle (3);
\draw (32.5,-27) node {$(0,0)$};
\draw [black] (32.5,-27) circle (2.4);
\draw [black] (44.8,-27) circle (3);
\draw (44.8,-27) node {$(1,1)$};
\draw [black] (23.7,-27) -- (29.5,-27);
\fill [black] (29.5,-27) -- (28.7,-26.5) -- (28.7,-27.5);
\draw [black] (34.818,-25.123) arc (118.34915:61.65085:8.07);
\fill [black] (42.48,-25.12) -- (42.02,-24.3) -- (41.54,-25.18);
\draw (38.65,-23.66) node [above] {$41$};
\draw [black] (42.382,-28.75) arc (-64.14001:-115.85999:8.557);
\fill [black] (34.92,-28.75) -- (35.42,-29.55) -- (35.86,-28.65);
\draw (38.65,-30.11) node [below] {$42$};
\end{tikzpicture}}

\subfloat[PD]{
\raisebox{12mm}{
\begin{tikzpicture}[scale=0.1]
\tiny
\tikzstyle{every node}+=[inner sep=0pt]
\draw [black] (32.5,-27) circle (3);
\draw (32.5,-27) node {$(0,0)$};
\draw [black] (32.5,-27) circle (2.4);
\draw [black] (44.8,-27) circle (3);
\draw (44.8,-27) node {$(1,1)$};
\draw [black] (23.7,-27) -- (29.5,-27);
\fill [black] (29.5,-27) -- (28.7,-26.5) -- (28.7,-27.5);
\draw [black] (34.818,-25.123) arc (118.34915:61.65085:8.07);
\fill [black] (42.48,-25.12) -- (42.02,-24.3) -- (41.54,-25.18);
\draw (38.65,-23.66) node [above] {$81$};
\draw [black] (42.382,-28.75) arc (-64.14001:-115.85999:8.557);
\fill [black] (34.92,-28.75) -- (35.42,-29.55) -- (35.86,-28.65);
\draw (38.65,-30.11) node [below] {$82$};
\end{tikzpicture}}}
~
\subfloat[Lathe]{
\begin{tikzpicture}[scale=0.1]
\tiny
\tikzstyle{every node}+=[inner sep=0pt]
\draw [black] (32.5,-27) circle (3);
\draw (32.5,-27) node {$(0,0)$};
\draw [black] (32.5,-27) circle (2.4);
\draw [black] (42.6,-18.2) circle (3);
\draw (42.6,-18.2) node {$(1,1)$};
\draw [black] (41.9,-35.7) circle (3);
\draw (41.9,-35.7) node {$(2,1)$};
\draw [black] (23.7,-27) -- (29.5,-27);
\fill [black] (29.5,-27) -- (28.7,-26.5) -- (28.7,-27.5);
\draw [black] (33.723,-24.268) arc (149.17489:112.95554:12.808);
\fill [black] (39.73,-19.04) -- (38.79,-18.89) -- (39.18,-19.81);
\draw (34.8,-20.68) node [above] {$51$};
\draw [black] (41.192,-20.843) arc (-33.79983:-64.06973:14.937);
\fill [black] (35.31,-25.97) -- (36.25,-26.07) -- (35.81,-25.17);
\draw (40.1,-24.29) node [below] {$52$};
\draw [black] (35.39,-27.769) arc (67.26662:27.16289:10.968);
\fill [black] (40.91,-32.88) -- (40.99,-31.94) -- (40.1,-32.39);
\draw (40.12,-29.35) node [above] {$53$};
\draw [black] (39.157,-34.496) arc (-119.30577:-146.26472:15.331);
\fill [black] (33.91,-29.64) -- (33.94,-30.58) -- (34.77,-30.03);
\draw (34.73,-32.86) node [below] {$54$};
\end{tikzpicture}}
~
\subfloat[$C_3$]{
\begin{tikzpicture}[scale=0.1]
\tiny
\tikzstyle{every node}+=[inner sep=0pt]
\draw [black] (32.5,-27) circle (3);
\draw (32.5,-27) node {$(0,0)$};
\draw [black] (32.5,-27) circle (2.4);
\draw [black] (42.6,-18.2) circle (3);
\draw (42.6,-18.2) node {$(1,1)$};
\draw [black] (41.9,-35.7) circle (3);
\draw (41.9,-35.7) node {$(2,1)$};
\draw [black] (23.7,-27) -- (29.5,-27);
\fill [black] (29.5,-27) -- (28.7,-26.5) -- (28.7,-27.5);
\draw [black] (33.723,-24.268) arc (149.17489:112.95554:12.808);
\fill [black] (39.73,-19.04) -- (38.79,-18.89) -- (39.18,-19.81);
\draw (34.8,-20.68) node [above] {$71$};
\draw [black] (41.192,-20.843) arc (-33.79983:-64.06973:14.937);
\fill [black] (35.31,-25.97) -- (36.25,-26.07) -- (35.81,-25.17);
\draw (40.1,-24.29) node [below] {$72$};
\draw [black] (35.39,-27.769) arc (67.26662:27.16289:10.968);
\fill [black] (40.91,-32.88) -- (40.99,-31.94) -- (40.1,-32.39);
\draw (40.12,-29.35) node [above] {$73$};
\draw [black] (39.157,-34.496) arc (-119.30577:-146.26472:15.331);
\fill [black] (33.91,-29.64) -- (33.94,-30.58) -- (34.77,-30.03);
\draw (34.73,-32.86) node [below] {$74$};
\end{tikzpicture}}

\subfloat[AM]{
\raisebox{13mm}{
\begin{tikzpicture}[scale=0.1]
\tiny
\tikzstyle{every node}+=[inner sep=0pt]
\draw [black] (32.5,-27) circle (3);
\draw (32.5,-27) node {$(0,0)$};
\draw [black] (32.5,-27) circle (2.4);
\draw [black] (44.3,-27) circle (3);
\draw (44.3,-27) node {$(1,1)$};
\draw [black] (54.4,-20.1) circle (3);
\draw (54.4,-20.1) node {$(2,1)$};
\draw [black] (54.4,-33.3) circle (3);
\draw (54.4,-33.3) node {$(3,1)$};
\draw [black] (23.7,-27) -- (29.5,-27);
\fill [black] (29.5,-27) -- (28.7,-26.5) -- (28.7,-27.5);
\draw [black] (35.5,-27) -- (41.3,-27);
\fill [black] (41.3,-27) -- (40.5,-26.5) -- (40.5,-27.5);
\draw (38.4,-27.5) node [below] {$61$};
\draw [black] (46.78,-25.31) -- (51.92,-21.79);
\fill [black] (51.92,-21.79) -- (50.98,-21.83) -- (51.54,-22.66);
\draw (50.85,-24.05) node [below] {$63$};
\draw [black] (46.85,-28.59) -- (51.85,-31.71);
\fill [black] (51.85,-31.71) -- (51.44,-30.86) -- (50.91,-31.71);
\draw (47.85,-30.65) node [below] {$65$};
\draw [black] (51.626,-34.432) arc (-72.98692:-139.11109:16.573);
\fill [black] (34.25,-29.43) -- (34.39,-30.36) -- (35.15,-29.71);
\draw (40.98,-35.09) node [below] {$66$};
\draw [black] (33.996,-24.405) arc (144.46403:70.51197:15.419);
\fill [black] (34,-24.41) -- (34.87,-24.04) -- (34.05,-23.46);
\draw (40.62,-18.1) node [above] {$64$};
\end{tikzpicture}}}
~
\subfloat[Robot]{
\begin{tikzpicture}[scale=0.1]
\tiny
\tikzstyle{every node}+=[inner sep=0pt]
\draw [black] (24.7,-32.6) circle (3);
\draw (24.7,-32.6) node {$(0,0)$};
\draw [black] (24.7,-32.6) circle (2.4);
\draw [black] (16.1,-16.9) circle (3);
\draw (16.1,-16.9) node {$(1,1)$};
\draw [black] (36.8,-17.6) circle (3);
\draw (36.8,-17.6) node {$(2,1)$};
\draw [black] (42.7,-32.6) circle (3);
\draw (42.7,-32.6) node {$(3,1)$};
\draw [black] (33.9,-48.2) circle (3);
\draw (33.9,-48.2) node {$(4,1)$};
\draw [black] (16.1,-48.2) circle (3);
\draw (16.1,-48.2) node {$(5,1)$};
\draw [black] (15.9,-32.6) -- (21.7,-32.6);
\fill [black] (21.7,-32.6) -- (20.9,-32.1) -- (20.9,-33.1);
\draw [black] (18.655,-18.463) arc (52.97882:4.44645:15.456);
\fill [black] (18.65,-18.46) -- (18.99,-19.34) -- (19.59,-18.55);
\draw (23.57,-22.19) node [right] {$31$};
\draw [black] (22.344,-30.747) arc (-132.62637:-169.94836:19.358);
\fill [black] (22.34,-30.75) -- (22.09,-29.84) -- (21.42,-30.57);
\draw (17.81,-26.99) node [left] {$32$};
\draw [black] (25.578,-29.734) arc (159.10446:123.11161:22.184);
\fill [black] (34.18,-19.06) -- (33.24,-19.08) -- (33.79,-19.92);
\draw (28.48,-22.29) node [left] {$33$};
\draw [black] (35.755,-20.41) arc (-23.70895:-54.07497:25.935);
\fill [black] (27.23,-30.98) -- (28.17,-30.92) -- (27.58,-30.11);
\draw (32.75,-27.69) node [right] {$34$};
\draw [black] (39.887,-33.637) arc (-73.66864:-106.33136:22.004);
\fill [black] (39.89,-33.64) -- (38.98,-33.38) -- (39.26,-34.34);
\draw (33.7,-35.02) node [below] {$30$};
\draw [black] (27.552,-31.676) arc (104.4624:75.5376:24.617);
\fill [black] (27.55,-31.68) -- (28.45,-31.96) -- (28.2,-30.99);
\draw (33.7,-30.4) node [above] {$39$};
\draw [black] (27.126,-34.36) arc (49.5845:11.47491:19.319);
\fill [black] (27.13,-34.36) -- (27.41,-35.26) -- (28.06,-34.5);
\draw (31.89,-38.01) node [right] {$37$};
\draw [black] (31.694,-46.169) arc (-135.89963:-163.04096:26.361);
\fill [black] (31.69,-46.17) -- (31.5,-45.25) -- (30.78,-45.94);
\draw (27.27,-42.46) node [left] {$38$};
\draw [black] (16.011,-45.206) arc (-184.00818:-233.72613:15.053);
\fill [black] (22.12,-34.12) -- (21.18,-34.19) -- (21.77,-35);
\draw (17.18,-37.8) node [left] {$35$};
\draw [black] (24.115,-35.541) arc (-14.72473:-43.00959:24.758);
\fill [black] (18.27,-46.14) -- (19.19,-45.89) -- (18.45,-45.21);
\draw (22.52,-42.4) node [right] {$36$};
\end{tikzpicture}}

\caption{Plants of the Flexible Manufacturing System}
\label{fig:FMS_A}
\end{figure}

The Flexible Manufacturing System (FMS) \citep{Queiroz2005} is composed of eight machines: three conveyors ($C_1$, $C_2$ and $C_3$), a mill, a lathe, a robot, a painting device (PD) and an assembly machine (AM). In each machine, the number of active tasks is represented in the state label so, for a state $(q, i)$, we have $f_{at}((q,i)) = i$. The machine models are shown in Figure~\ref{fig:FMS_A}. The initial state of each machine is an idle state, with no active tasks, and the other states have one active task each. Controllable events are represented by odd numbers and the uncontrollable ones are represented by even numbers.

The FMS produces two kinds of products, a Product A and a Product B. Both products share the same base, given by the following sequence of pairs (the controllable events must obey the order in the sequence, but the uncontrollable events may occur in any order allowed in the supervisor):
\begin{align*}
b = \{(11,12),(31,32),(41,42),(35,36),(61)\}.
\end{align*}
To produce the pin of a Product A, the pairs to be executed are:
\begin{align*}
p_a = \{(21,22),(33,34),(51,52),(37,38),(63,64)\}.
\end{align*}
Finally, to produce the pin of a Product B, the pairs are:
\begin{align*}
p_b = \{ & (21,22),(33,34),(53,54),(39,30),(71,72),\\ & (81,82),(73,74),(65,66)\}.
\end{align*}
The monolithic supervisor of the FMS, using the Supervisory Control Theory, has $45,504$ states and $200,124$ transitions. 

In the tests, a batch that produces one Product A and one Product B is considered a batch of size one, so, a batch of size $N$ produces $N$ Products A and $N$ Products B. Each pair of products (one A and one B) is represented by a sequence of 44 events, so a batch of size $N$ is represented by $44 \times N$ events.

In order to use time, we have to define the time interval between related events, as shown in Table~\ref{tab:times}. The FMS has a peculiarity, event 61 does not have an uncontrollable counterpart. Then, the occurrence of 63 and 65 is at least 15 time units after 61.


\begin{table}[H]
\caption{Time interval between related events, (in time units - t.u.)}
\centering
\begin{tabular}{|l|c|c|c|}
\hline
Plants &\begin{tabular}[c]{@{}c@{}} Controllable\\ Events\end{tabular} & \begin{tabular}[c]{@{}c@{}}Uncontrollable \\ Events\end{tabular} & \begin{tabular}[c]{@{}c@{}}Operation\\ Time $[t.u.]$\end{tabular} \\ \hline
$C_1$ & $11$ & $12$ & $25$ \\ \hline
$C_2$ & $21$ & $22$ & $25$ \\ \hline
& $31$ & $32$ & $21$ \\ 
& $33$ & $34$ & $19$ \\ 
Robot& $35$ & $36$ & $16$ \\ 
& $37$ & $38$ & $24$ \\ 
& $39$ & $30$ & $20$ \\ \hline
Mill & $41$ & $42$ & $30$ \\ \hline
\multirow{2}{*}{Lathe}& $51$ & $52$ & $38$ \\ 
& $53$ & $54$ & $32$ \\ \hline
& $61$ & -  & $15$ \\ 
AM& $63$ & $64$ & $26$ \\ 
& $65$ & $66$ & $26$ \\ \hline
\multirow{2}{*}{$C_3$}& $71$ & $72$ & $25$ \\ 
& $73$ & $74$ & $25$ \\ \hline
PD & $81$ & $82$ & $24$ \\ \hline
\end{tabular}

\label{tab:times}
\end{table}

\subsection{Results}
\begin{figure}[ht!]
\centering
{\includegraphics[width=0.9\columnwidth]{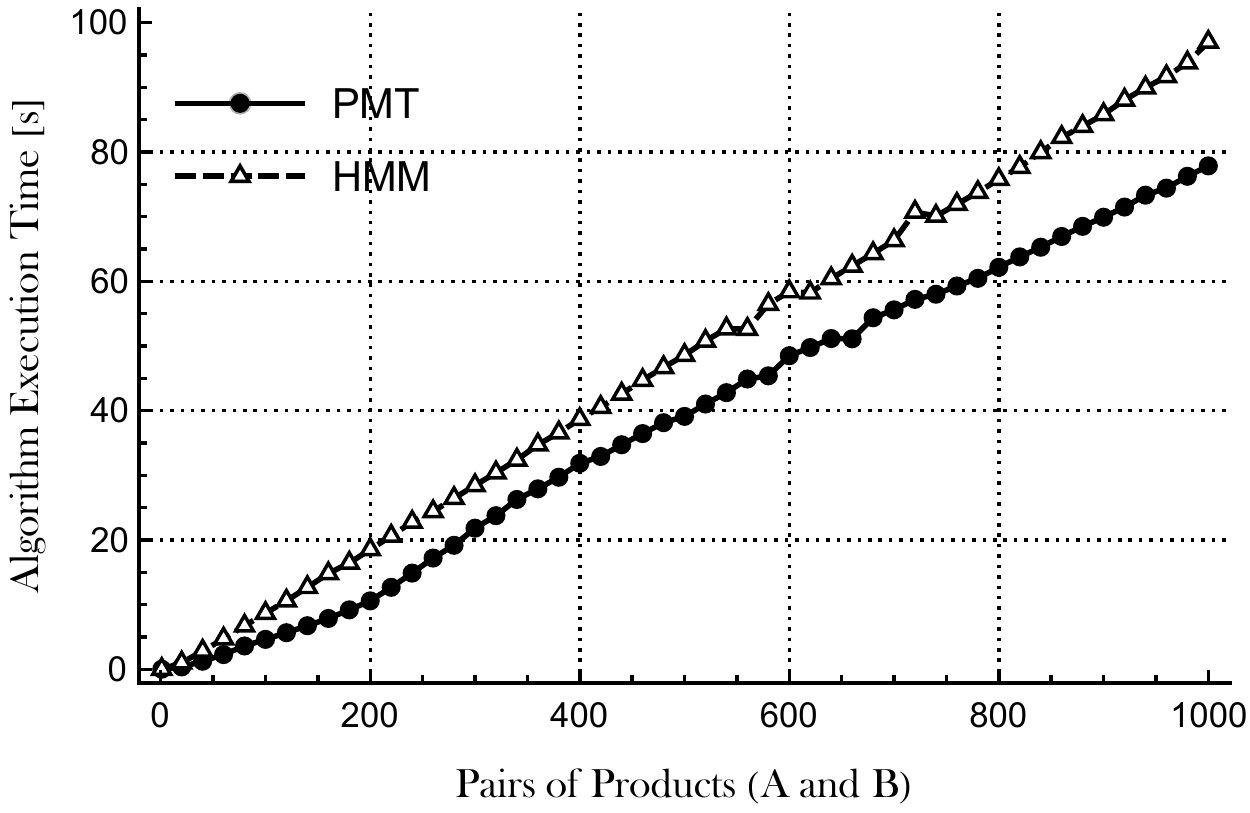}}
\\
\caption{Algorithms Execution Time}
\label{fig:tRes}

\end{figure}

Two algorithms were applied: the Parallelism Maximization with Time Restrictions (PMT) and the Heuristic Mak\-kes\-pan Minimization (HMM).
Each algorithm was executed once for batch sizes of one pair of products to 1000 pairs of products and the results of PMT and HMM are shown in Figure~\ref{fig:tRes} and Tables \ref{tab:tRes} and \ref{tab:tResP}. The computations were performed in a computer with CPU \textit{Intel Xeon E5-2667} 2.90~GHz and 64~GB of RAM memory.

As we can see in Fig.~\ref{fig:tRes}, the fastest algorithm is the Parallelism Maximization with Time Restrictions (PMT), finding a sequence to produce 1000 pairs of products in less than 80 seconds. The Heuristic Makespan Minimization (HMM) is slower, but also has a good execution time, around 100 seconds for $N=1000$.

Both algorithms, PMT and HMM, give good results regarding the makespan, but in all situations HMM gives a smaller makespan, at the cost of a higher execution time, as can be seen in Table \ref{tab:tRes}.
If we use as a baseline the result obtained in \citep{Robi2018}, that applies Model Checking to the same problem, for $N \leq 15$, we see that HMM hits the lower bound in all situations in which the optimal result is known (batches for which the Model Checking resulted, up to $N=15$). The authors in \citep{Robi2018} also provide a formula to calculate the optimal makespan for any batch size,
\begin{equation}\label{eq:formula}
T(N) = 157N + 81.
\end{equation}

With HMM, we are able to find a sequence that has the makespan value predicted by (\ref{eq:formula}) even when there is no exact procedure for finding it.

The makespan in \citep{Robi2018} is shown in Table~\ref{tab:tRes} as the optimal value, referred to as MC (from Model Checking). It is important to notice that the ability to find the optimal makespan for any size of batch does not correspond to finding the sequence that will provide such a makespan. Using model checking the sequence could be found only to batches up to $N=15$, and the algorithm took over one hour to find the result (against 0.56s that the HMM took to find the same result). To sizes greater than $N=15$, the execution ran out of memory. In Table~\ref{tab:tRes} the time execution of the model checking is presented, up to $N=15$.

\begin{table*}[ht]
\centering
\caption{Makespan obtained using the algorithms PMT, HMM and Model Checking (MC) \citep{Robi2018} and their corresponding execution time (MC was executed on a PC with a 2.8~GHz CPU and 16~GB of RAM).}
\label{tab:tRes}
\begin{tabular}{|r|r|r|r|r|r|r|}
\hline
\multicolumn{1}{|c|}{Batch} & \multicolumn{2}{c|}{PMT} & \multicolumn{2}{c|}{HMM} &\multicolumn{2}{c|}{MC} \\ \cline{2-7} 
\multicolumn{1}{|c|}{Size} & \multicolumn{1}{c|}{Makesp.} & \multicolumn{1}{c|}{Exec.T.} &  \multicolumn{1}{c|}{Makesp.} & \multicolumn{1}{c|}{Exec.T.} & \multicolumn{1}{c|}{Makesp.} & \multicolumn{1}{c|}{Exec.T.}\\ \hline
 $1$      &  $\textbf{238}$     &  $0.025$ sec   & $\textbf{238}$     & $0.026$ sec  & $\textbf{238}$    & $0.6$ min \\ \hline
 $5$      &  $878$     &  $0.054$ sec   & $\textbf{866}$    &  $0.02$ sec & $\textbf{866}$     & $10.2$ min  \\ \hline
 $10$    &  $1,663$   &  $0.22$ sec   & $\textbf{1,651}$   &  $0.32$ sec  & $\textbf{1,651}$  & $30.2$ min   \\ \hline
 $15$    &  $2,448$   &  $0.51$ sec   & $\textbf{2,436}$   &  $0.56$ sec  & $\textbf{2,436}$   & $61.7$ min   \\ \hline
 $50$     & $7,943$   &  $1.80$ sec   & $\textbf{7,931}$   &  $4.25$ sec  & $\textbf{7,931}$   & $-$  \\ \hline
 $100$    &  $15,793$  &  $3.39$ sec   & $\textbf{15,781}$  &  $8.59$ sec & $\textbf{15,781}$  & $-$  \\ \hline
 $500$    &  $78,593$  & $16.13$ sec   & $\textbf{78,581}$  & $48.74$ sec & $\textbf{78,581}$  & $-$  \\ \hline
 $750$   &  $117,843$ & $24.40$ sec   & $\textbf{117,831}$ & $74.49$ sec & $\textbf{117,831}$ & $-$  \\ \hline
 $1000$   &  $157,093$ & $32.10$ sec   & $\textbf{157,081}$ & $97.26$ sec & $\textbf{157,081}$ & $-$  \\ \hline
\end{tabular}
\end{table*}

The analysis of the cumulative parallelism shows that the PMT gives a slightly bigger parallelism, and a slightly smaller makespan, as shown in Table~\ref{tab:tResP}. Although parallelism is a good indicator of performance, the best sequence in terms of makespan may not have the biggest parallelism. Since checking the accumulated parallelism is computationally cheaper, a sequence that maximizes parallelism may be used as a starting point to other algorithms.

\begin{table}[ht]
\centering
\caption{Cumulative Parallelism obtained using the algorithms PMT and HMM}
\label{tab:tResP}
\begin{tabular}{|r|r|r|}
\hline
\multicolumn{1}{|c|}{Batch} & \multicolumn{2}{c|}{Parallelism} \\ \cline{2-3} 
\multicolumn{1}{|c|}{size} & \multicolumn{1}{c|}{PMT} & \multicolumn{1}{c|}{HMM} \\ \hline
 $1$    &  $\textbf{93}$     & $\textbf{93}$     \\ \hline
 $5$    &  $\textbf{713}$    & $635$      \\ \hline
 $10$   & $\textbf{1,488}$   & $1,315$   \\ \hline
 $15$   & $\textbf{2,263}$   & $1,995$    \\ \hline
 $50$   & $\textbf{7,688}$   & $6,755$  \\ \hline
 $100$  & $\textbf{15,438}$  & $13,555$   \\ \hline
 $500$  & $\textbf{77,438}$  & $67,955$ \\ \hline
 $750$  & $\textbf{116,188}$ & $101,955$ \\ \hline
 $1000$ & $\textbf{154,938}$ & $135,955$  \\ \hline
\end{tabular}
\end{table}

\subsection{Robustness to Model Uncertainty}


The durations of each operation were considered known and deterministic, as other approaches from the literature such as those presented in \citep{framinan2019,1678411,6042515,5771983,Su2016, Ware2017,Su2017}. In real life industrial applications, however, it is usually the case that such durations may vary a bit from one execution to another. When the operation times used in the planning problem match the duration of the operations in real life, then the uncontrollable events occur in the positions that the algorithms predict. On the other hand, when the durations are different, if the complete sequence is implemented, infeasibility would be generated, caused by the specified order for the uncontrollable events. It makes sense, however, to consider that only the sequence of controllable events is implemented in the control system (uncontrollable events are responses of the system and should not be fixed by the control system). Because of the controllability property of the Supervisory Control, the sequence of controllable events of the solution, interleaved with uncontrollable events in any order that can be generated by the plant, is feasible in the controlled system. To evaluate the performance of the solutions under disturbances, we use a solution provided by the algorithm, remove the uncontrollable events, implement disturbances in the durations of the operations and evaluate the makespan of the sequence.


Specifically, we simulated two sequences that produce 100 pairs of products, one generated by the PMT algorithm and other generated by the HMM algorithm. For both sequences, generated using the operation times of Table~\ref{tab:times}, we removed the uncontrollable events and simulated the system with normal distributed random times, with mean equal to the original time and the standard deviation ($\sigma$) was varied from $0$ (deterministic case) to $5$.

\begin{figure}[ht!]
\centering
\subfloat[Makespan] {\includegraphics[width=0.9\columnwidth]{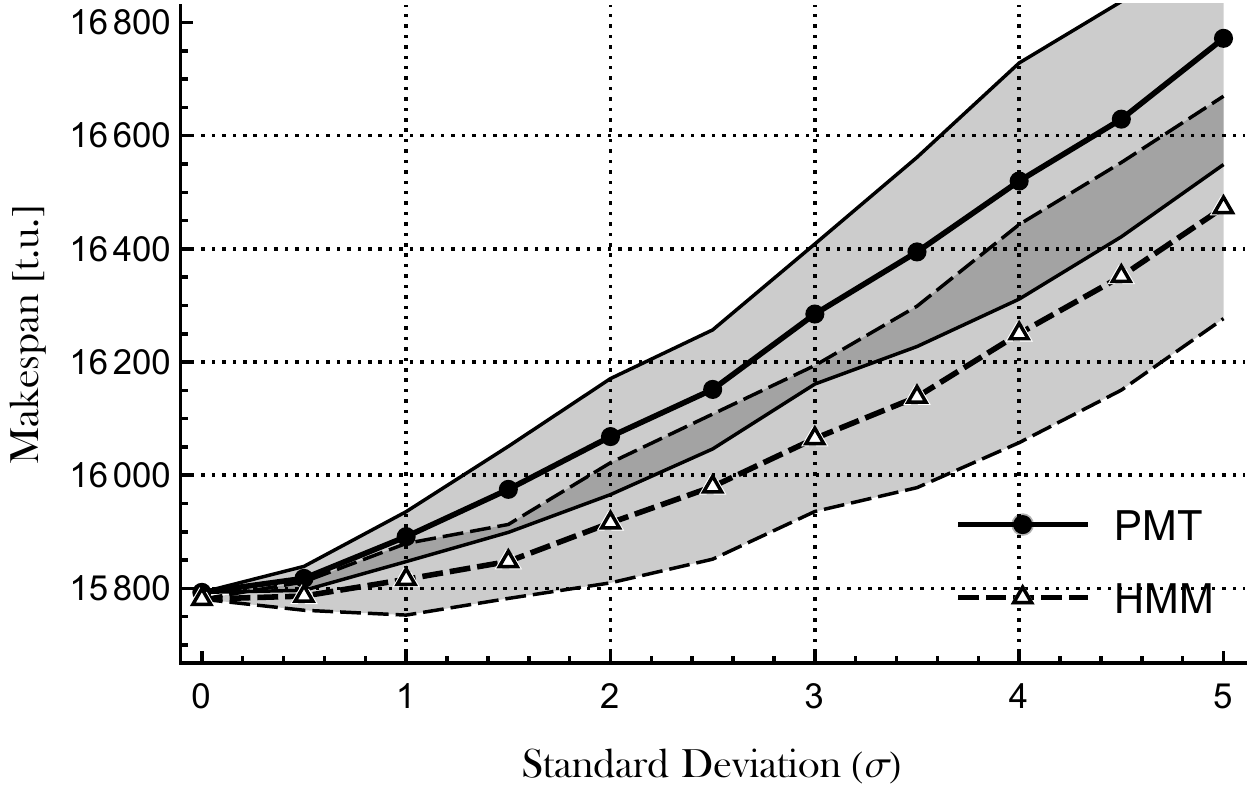}}
\\
\subfloat[Cumulative Parallelism] {\includegraphics[width=0.9\columnwidth]{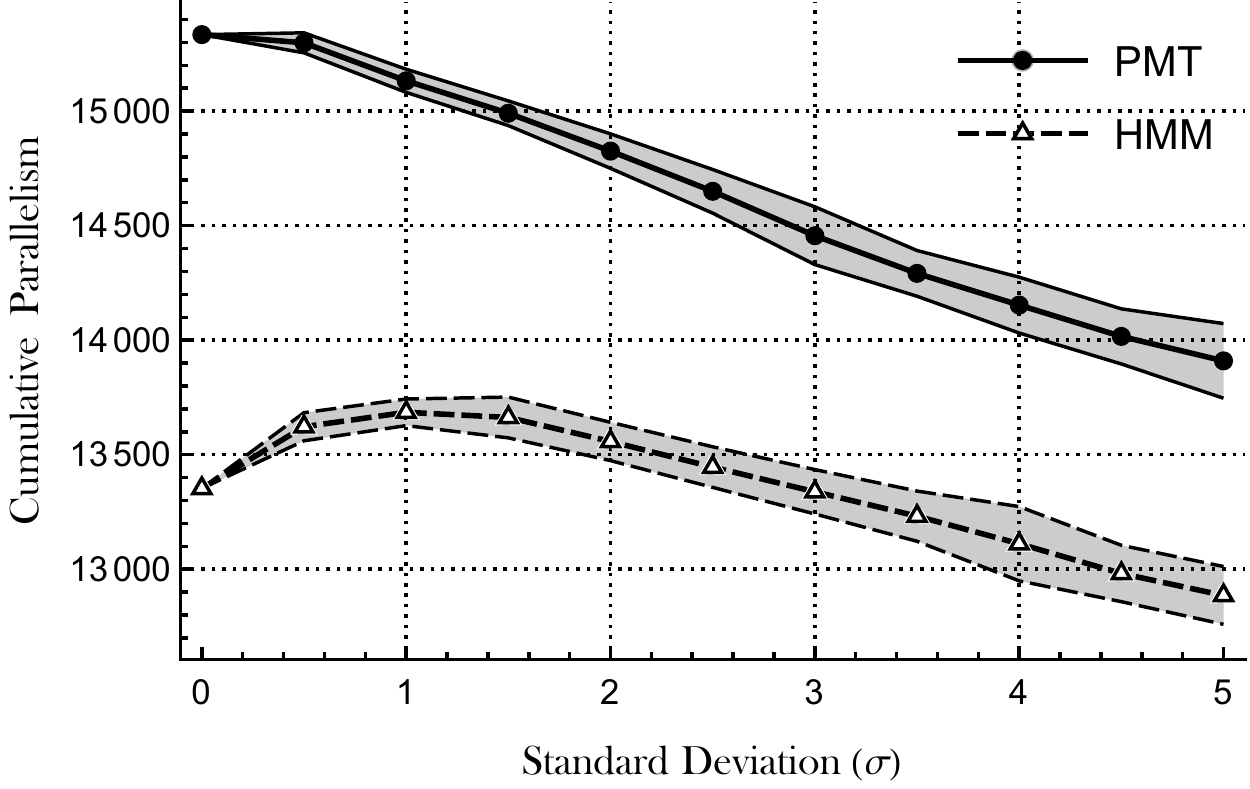}}

\caption{Results of algorithms with disturbances in the machine times}
\label{fig:rob}

\end{figure}

For each standard deviation the simulation was executed 30 times and the mean ($\mu_r$) and the standard deviation ($\sigma_r$) of the results were recorded. Figure~\ref{fig:rob} shows the result of the simulations: the mean of the results forms the central line and the shaded area around each line is a confidence interval $(\mu_r \pm 2 \sigma_r)$. As can be seen, as the standard deviation is increased, the results become worse, which is expected given that the simulation times tends to deviate more from those obtained through the application of the algorithm. Although the HMM solution presents a smaller mean makespan, the confidence intervals of the solutions obtained by both algorithms overlap, indicating that, in a real life application with variability in the durations, both algorithms have similar performance.  

\section{Conclusion}\label{sec:conc}

This paper presents two efficient algorithms based on the idea of maximizing the parallelism among equipment to minimize makespan. Both algorithms take heuristic steps, that allow finding the solution even for very large batches of products. Both procedures belong to the class of heuristics and the main difference among then is the fact that PMT is a parallelism-oriented heuristic with polynomial complexity and HMM is a time-oriented heuristic with a non-polynomial complexity. The main concepts are illustrated in a small example and then a case study, previously solved with Model Checking, is used to illustrate the efficiency of the algorithms. 

The Parallelism Maximization with Time Restrictions (PMT) is a polynomial algorithm, slightly faster than the Heuristic Makespan Minimization, with a good compromise between execution time and makespan results. The Heuristic Makespan Minimization (HMM) is a non-polynomial algorithm with a heuristic step that allows an enormous reduction on the branching factor of an exact algorithm. With this reduction, HMM hits the optimal makespan value for all batch sizes for which it is known (for the case study), and the execution time is much smaller. Additionally, HMM allowed to find a solution with the optimal makespan predicted in \citep{Robi2018} for batches up to 1000 products.

Finally, the results show that using the parallelism is a good strategy to solve scheduling problems. In fact, our results show that increasing parallelism while respecting time constraints is a good way to increase production performance, even though there are multiple sequences with the same ma\-ke\-span but different levels of parallelism.

As future work, we intend to evaluate the performance of parallelism maximization as a indirect criterion for other optimization problems. 

\section*{Acknowledgment}

This work has been supported by the Brazilian agencies CAPES, CNPq and Fapemig.

\appendix
\section{Appendix}

This example illustrates the execution of Algorithm \ref{alg:PMTa}, over the system in Example \ref{ex:Sm1}. 
\begin{example} Consider the small factory, presented in Example \ref{ex:Sm31}. The execution of Algorithm \ref{alg:PMTa} is presented in Fig.~\ref{fig:sim2}. In this approach, the duration of the operation of the machines is part of the optimization, so a time interval of $10\;t.u.$ is considered between events $a_1$ and $b_1$ and $5\;t.u.$ between events $a_2$ and $b_2$.
\begin{figure*}[htbp]
    \centering
    \subfloat[IIE0:\protect\\ $f_T(path(IIE0) b_1)=\infty$,\protect\\ $f_T(path(IIE0) b_2)=\infty$,\protect\\ $F_{at}=0$] {\includegraphics[width=0.3\textwidth]{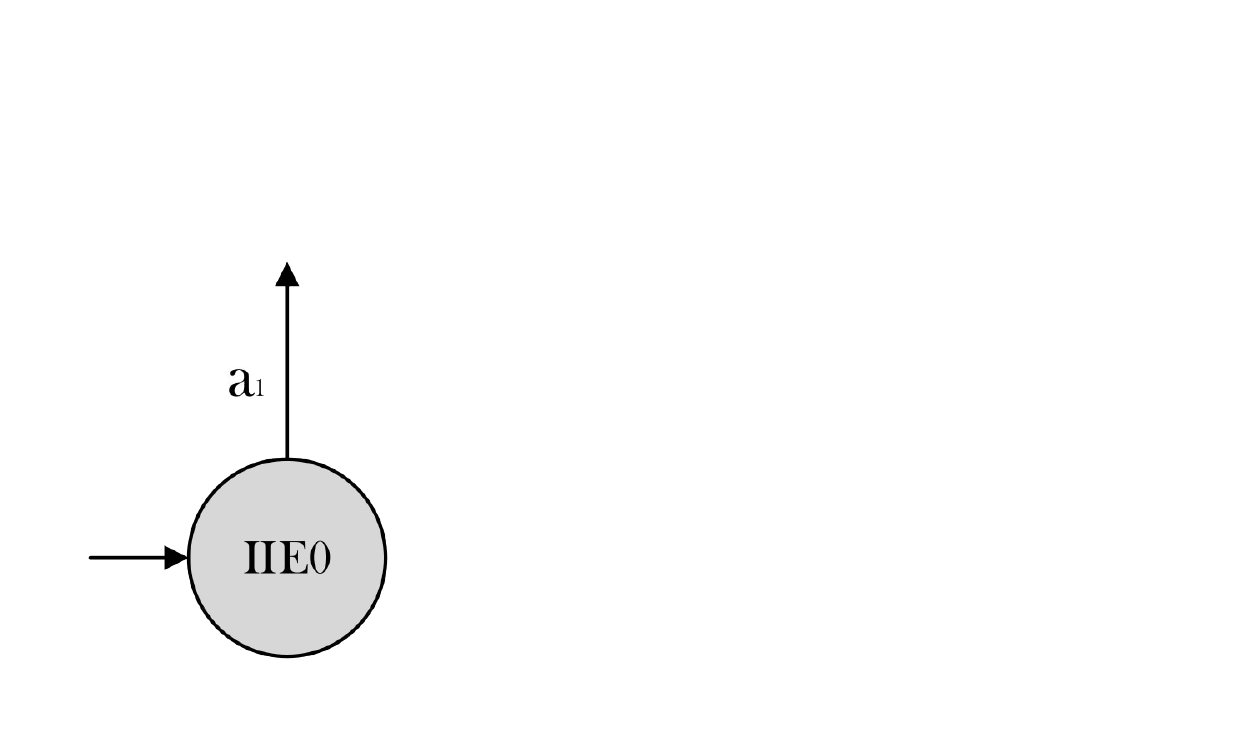}}~
    \subfloat[WIE1:\protect\\ $f_T(path(WIE1) b_1)=10$,\protect\\ $f_T(path(WIE1) b_2)=\infty$,\protect\\ $F_{at}=1$] {\includegraphics[width=0.3\textwidth]{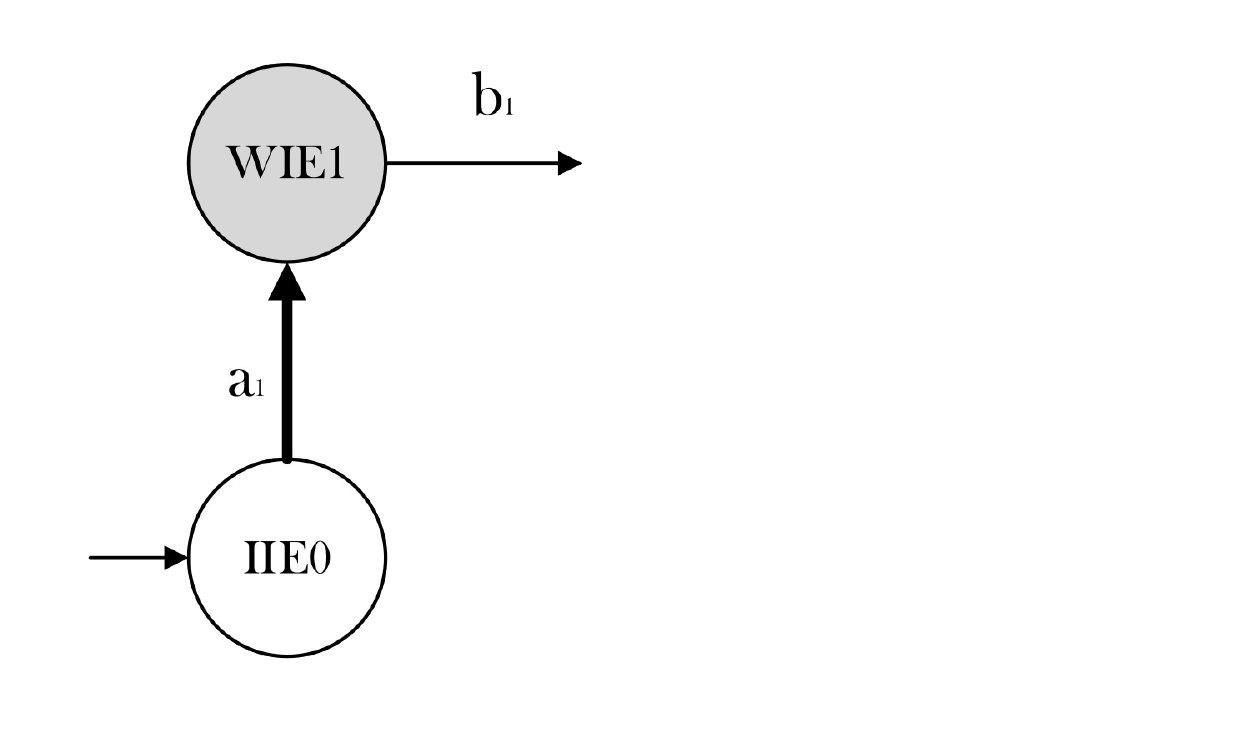}}~
    \subfloat[IIF2:\protect\\ $f_T(path(IIF2) b_1)=\infty$,\protect\\ $f_T(path(IIF2) b_2)=\infty$,\protect\\ $F_{at}=1$] {\includegraphics[width=0.3\textwidth]{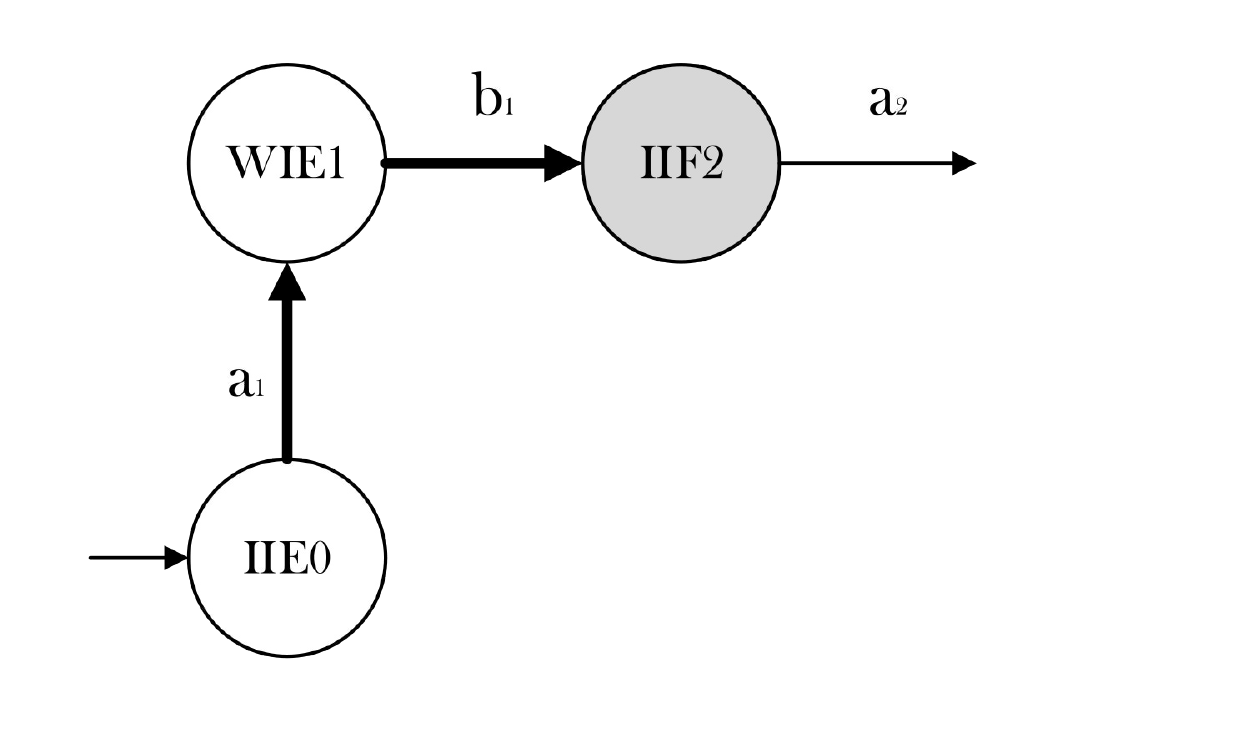}}
  
    \subfloat[IWE3:\protect\\ $f_T(path(IWE3) b_1)=\infty$,\protect\\ $f_T(path(IWE3) b_2)=5$,\protect\\ $F_{at}=2$] {\includegraphics[width=0.3\textwidth]{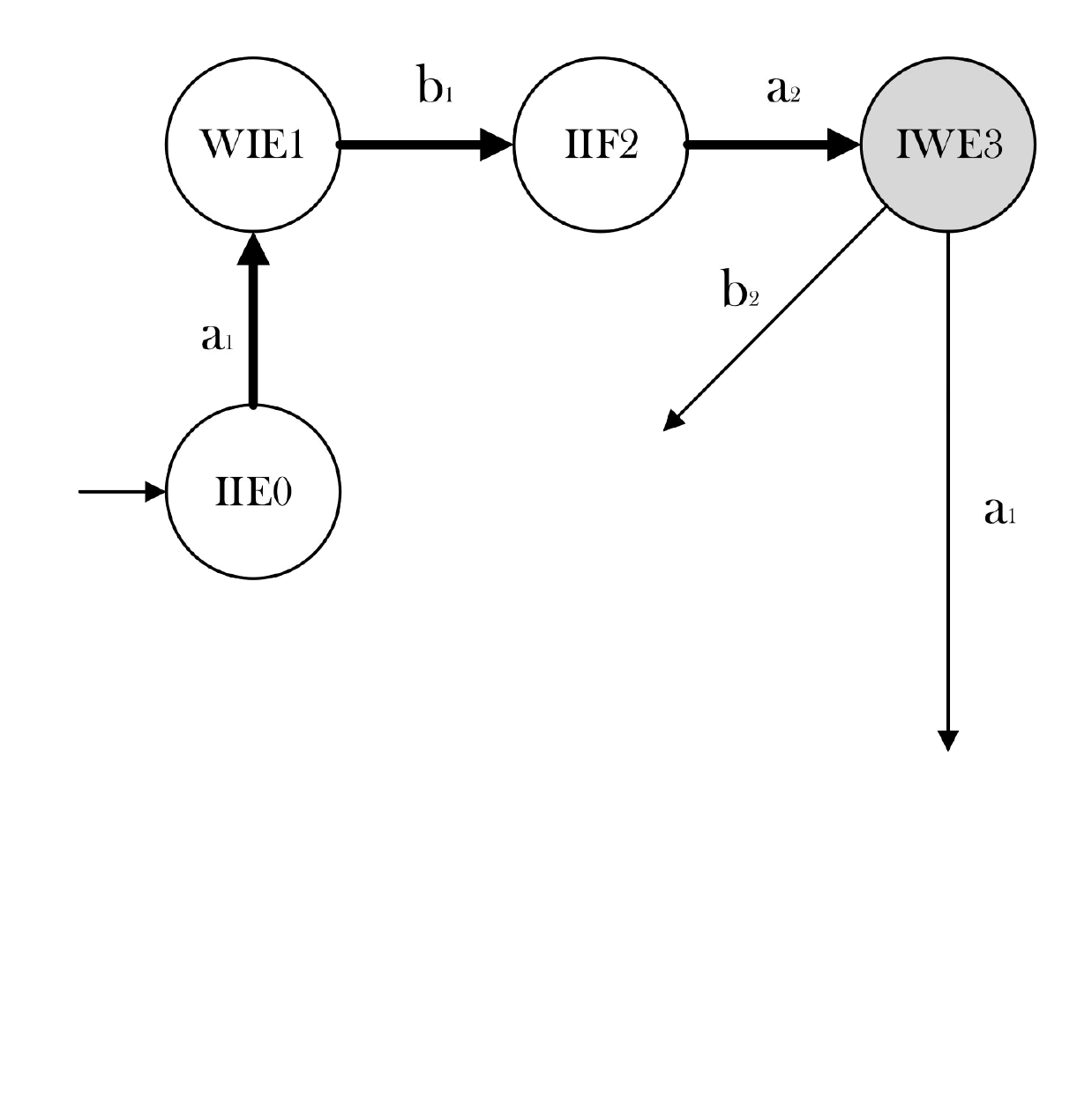}}~
    \subfloat[WWE4:\protect\\ $f_T(path(WWE4) b_1)=10$,\protect\\ $f_T(path(WWE4) b_2)=5$,\protect\\ $F_{at}=4$;] {\includegraphics[width=0.3\textwidth]{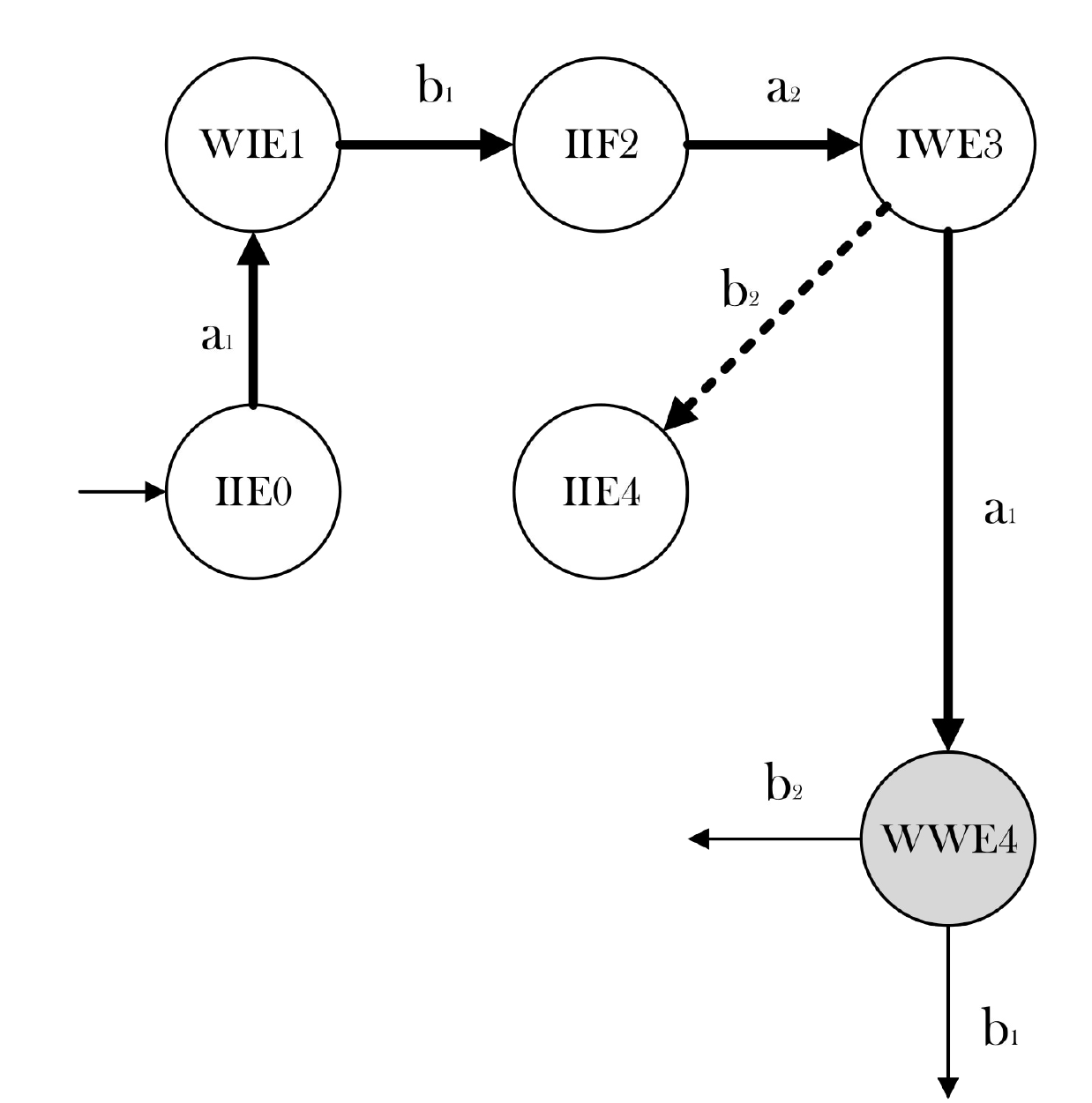}}~ 
    \subfloat[WIE5:\protect\\ $f_T(path(WIE5) b_1)=5$,\protect\\ $f_T(path(WIE5) b_2)=\infty$,\protect\\ $F_{at}=5$] {\includegraphics[width=0.3\textwidth]{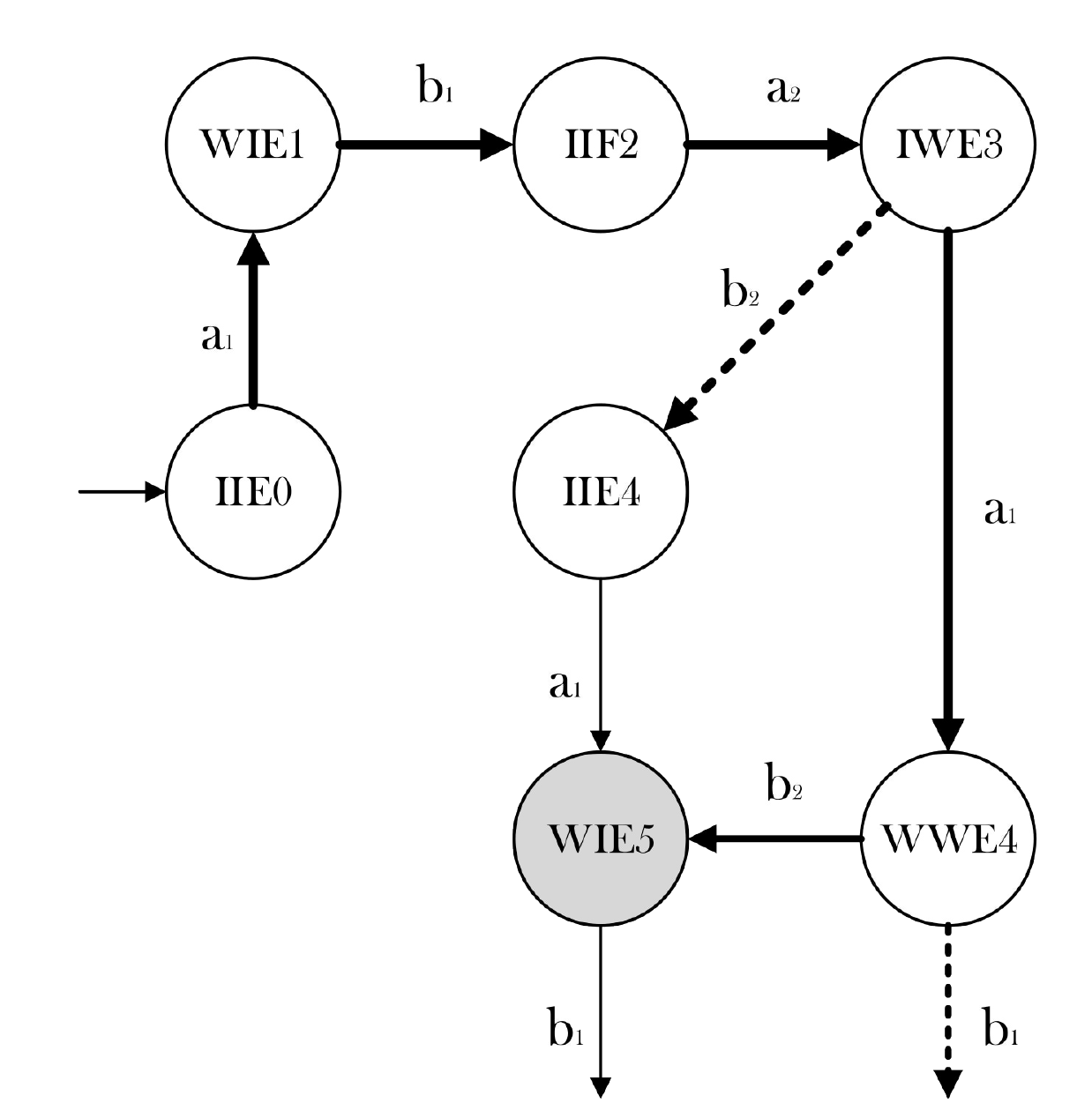}}
    
    \subfloat[IIF6:\protect\\ $f_T(path(IIF6) b_1)=\infty$,\protect\\ $f_T(path(IIF6) b_2)=\infty$,\protect\\ $F_{at}=5$] {\includegraphics[width=0.3\textwidth]{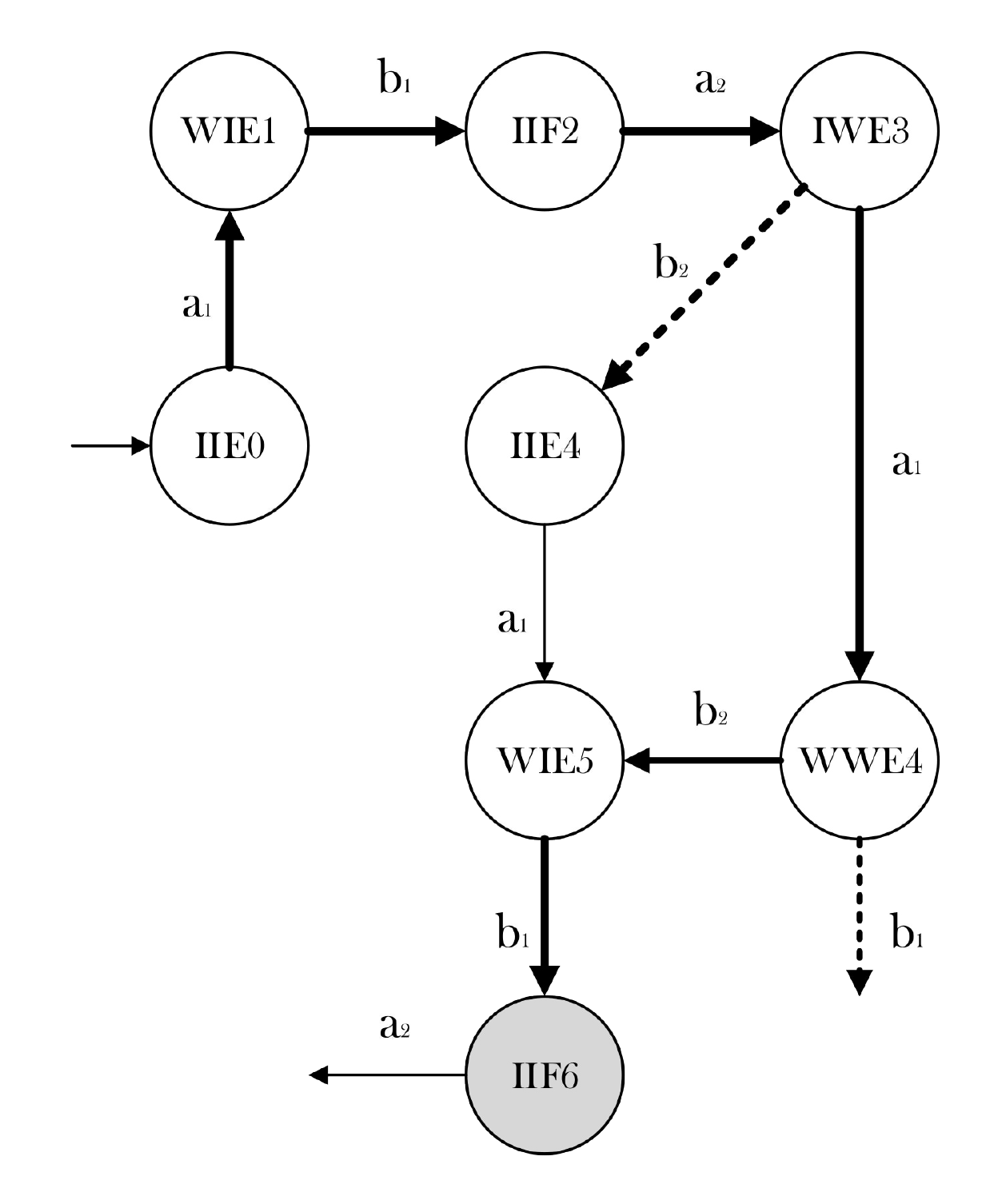}}~
    \subfloat[IWE7:\protect\\ $f_T(path(IWE7) b_1)=\infty$,\protect\\ $f_T(path(IWE7) b_2)=5$,\protect\\ $F_{at}=6$] {\includegraphics[width=0.3\textwidth]{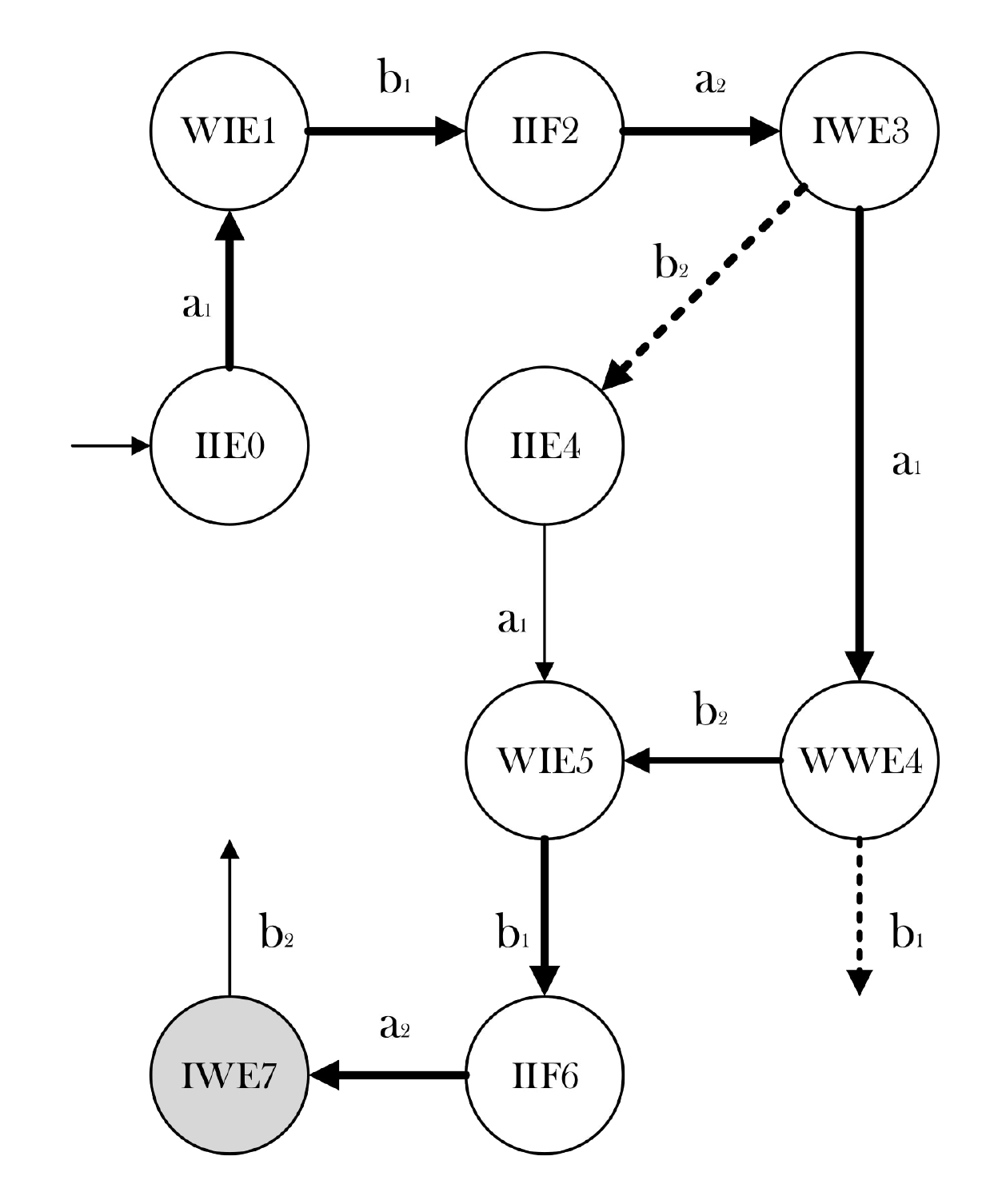}}~
    \subfloat[IIE8:\protect\\ $f_T(path(IIE8) b_1)=\infty$,\protect\\ $f_T(path(IIE8) b_2)=\infty$,\protect\\ $F_{at}=6$] {\includegraphics[width=0.3\textwidth]{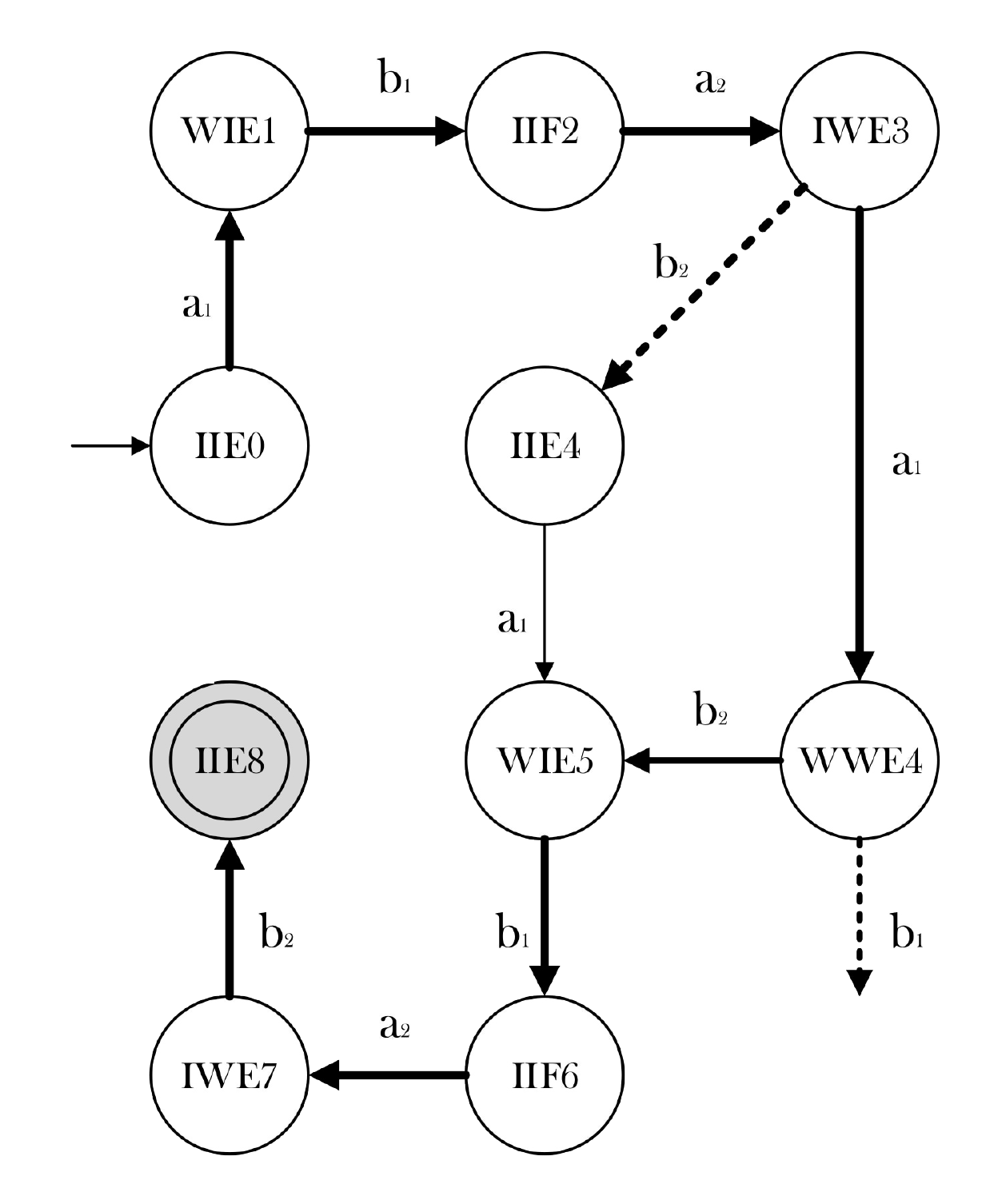}}
    \caption{Example of the execution of the Parallelism Maximization with Time Restrictions Algorithm for the Small Factory Problem producing two products.}
    \label{fig:sim2}
\end{figure*}
In each step, starting in the initial state, the algorithm travels the closed loop system accumulating the number of tasks in the path and updating a schedule with the time until the occurrence on each uncontrollable event. For instance, at the initial step, Fig.\ref{fig:sim2} (a), $T(b_1)=T(b_2)=\infty$ since such events are now allowed in the supervisor. When state $WIE1$ is reached, $T(b_1)=10$, since $a_1$ has occurred, and $T(b_2)=\infty$, in Figures~\ref{fig:sim2} (a) (b) (c) there is only one event to execute but when the state $IWE3$ is reached (Figure~\ref{fig:sim2} (d)) the events $b_2$ and $a_1$ can be executed, but only state WWE4, reached by executing event $a_1$, is visited because the algorithm executes controllable events when they are possible (heuristic step). In state WWE4 (Figure~\ref{fig:sim2} (e)), events $b_1$ and $b_2$ are logically possible to occur but the occurrence of $b_2$  is temporally infeasible. In Figures~\ref{fig:sim2} (f) (g) (h) (i) the algorithm has only one path to follow, reaching the final state $IIE8$. The algorithm finishes when it reaches a marked state, after executing 8 events. The resulting sequence in this example is $s^* = a_1\,b_1\,a_2\,a_1\,b_2\,b_1\,a_2\,b_2$, the parallelism $F_{at}(s^*)= 6$ and the makespan of the sequence is $f_T(s^*)=25\;t.u.$.
\end{example}

\printcredits

\bibliographystyle{cas-model2-names}

\bibliography{biblio}

\end{document}